\documentclass[aps,prl,reprint]{revtex4-1}
\pdfoutput=1 

\usepackage{blindtext}
\usepackage{hyperref}
\usepackage{bbold}
\usepackage{mathtools, cases}
\usepackage[T1]{fontenc} 
\usepackage{float}

\newcommand{\bk}[1]{\boldsymbol{k}_{#1}}
\newcommand{\bq}[1]{\boldsymbol{q}_{#1}}
\newcommand{\fnl}[1]{f_\text{NL}^{\text{#1}}}

\newcommand{\qv}{\mathbf{q}}

\def\kmin{k_\mathrm{min}}
\def\kmax{k_\mathrm{max}}

\newcommand{\RUG}{Kapteyn Astronomical Institute, University of Groningen, 9700~AV Groningen, The Netherlands}
\newcommand{\VSI}{Van Swinderen Institute for Particle Physics and Gravity,\\ University of Groningen,
Nijenborgh 4, 9747 AG Groningen, The Netherlands}
\newcommand{\SISSA}{International School for Advanced Studies~(SISSA), 34136~Trieste, Italy}
\newcommand{\IFPU}{Institute for Fundamental Physics of the Universe, via Beirut 2, 34151 Trieste, Italy}
\newcommand{\INAF}{INAF, Osservatorio Astronomico di Trieste, via Tiepolo 11, 34143 Trieste, Italy}
\newcommand{\INFN}{INFN, Sezione di Trieste, Via Valerio 2, 34127 Trieste, Italy}

\begin{document}
\title{Primordial non-Gaussianity and non-Gaussian Covariance}

\author{Thomas Fl\"oss}
\email{t.s.floss@rug.nl}
\affiliation{\VSI,\RUG}

\author{Matteo Biagetti}
\affiliation{\IFPU,\SISSA,\INAF,\INFN}

\author{P. Daniel Meerburg}
\affiliation{\VSI}

\begin{abstract}
In the pursuit of primordial non-Gaussianities, we hope to access smaller scales across larger comoving volumes. At low redshift, the search for primordial non-Gaussianities is hindered by gravitational collapse, to which we often associate a scale $k_{\rm NL}$. Beyond these scales, it will be hard to reconstruct the modes sensitive to the primordial distribution. When forecasting future constraints on the amplitude of primordial non-Gaussianity, $\fnl{}$, off-diagonal components are usually neglected in the covariance because these are small compared to the diagonal. We show that the induced non-Gaussian off-diagonal components in the covariance degrade forecast constraints on primordial non-Gaussianity, even when all modes are well within what is usually considered the linear regime. As a testing ground, we examine the effects of these off-diagonal components on the constraining power of the matter bispectrum on $\fnl{}$ as a function of $\kmax{}$ and redshift, confirming our results against N-body simulations out to redshift $z=10$. We then consider these effects on the hydrogen bispectrum as observed from a PUMA-like 21-cm intensity mapping survey at redshifts $2<z<6$ and show that not including off-diagonal covariance over-predicts the constraining power on $f_{\rm NL}$ by up to a factor of $5$.  For future surveys targeting even higher redshifts, such as Cosmic Dawn and the Dark Ages, which are considered ultimate surveys for primordial non-Gaussianity, we predict that non-Gaussian covariance would severely limit prospects to constrain $\fnl{}$ from the bispectrum. 

\end{abstract}

\maketitle

\section{Introduction}
Over the past few decades, inflation has been established as the leading paradigm for describing the early universe. It proposes a period of rapidly accelerated expansion during the first fraction of a second after the universe came to be \cite{Guth:1980zm,Linde:1981mu,Albrecht:1982wi}. At the classical level such an expansion can explain why the universe looks nearly identical in every direction (i.e. is homogeneous and isotropic), while at the quantum level it gives rise to the tiny density fluctuations that we observe in the cosmic microwave background radiation (CMB), which eventually grow into the large scale structure of the universe (LSS). By precisely mapping the anisotropies in the CMB, we have determined the fluctuations to be very close to Gaussian distributed, which matches the predictions of even the simplest theories of inflation \cite{Akrami:2018odb}. However, in order to sift through the vast landscape of consistent inflationary theories we are required to look beyond such general predictions. One avenue to discriminate theories of inflation, is through the study of primordial non-Gaussianities (pnGs) (see \cite{Meerburg:2019qqi} and references therein). Signatures of pnG would appear as non-zero higher $n$-point functions of the initial conditions, where the $3$-point function, the so-called \emph{bispectrum}, is generally the most sensitive. A measurement of pnGs can tell us a great deal about the dynamics driving the expansion (see \cite{Achucarro:2022qrl} for a recent overview). Furthermore, particles (fields) present in the primordial universe leave their unique imprint in the distribution of fluctuations through pnGs, effectively making inflation a particle collider at the highest conceivable energy scale \cite{Arkani-Hamed:2015bza}. Hence, a detailed study of primordial non-Gaussianity is imperative in order to advance our understanding of the universe as a whole. 

While the most stringent constraints on pnGs are derived from measurements of the CMB bispectrum, future CMB experiments will be limited by its two-dimensional nature and damping of primary fluctuations. In our search for signatures of pnGs we are therefore required to look for alternative probes. Surveys of the large scale structure of the universe provide us with a huge observable volume all the way into the cosmological Dark Ages, by mapping the distribution of galaxies and neutral hydrogen. While the anisotropies in the CMB are pristine (i.e. linearly related to the primordial fluctuations), the density field has since evolved. Gravity, being intrinsically non-linear, breaks the linear relation between density fluctuations and primordial initial conditions, giving rise to a number of complications. Firstly, even if the primordial fluctuations are purely Gaussian, the non-linear gravitational evolution introduces \emph{secondary} non-Gaussianities (snGs), typically many orders of magnitude stronger than any primordial signal. Thus, an accurate modelling of snG is required in order to properly extract information about pnG. Furthermore, snGs introduce non-Gaussian covariance in the measurements, reducing the amount of \emph{unique} information present in the data. Although the impact of non-Gaussian covariance has been appreciated at low redshifts \cite{Takahashi:2009ty,Chan:2016ehg,Chan:2017fiv,Wadekar:2019rdu,Barreira:2019icq,Gualdi:2020ymf,Oddo:2021iwq,Barreira:2021ueb,Biagetti:2021tua,Rizzo:2022lmh}, its relevance for high redshift surveys has typically been neglected \cite{Munoz:2015eqa,Chen:2016zuu,Meerburg:2016zdz,Karagiannis:2019jjx,Floss:2022grj,Yamauchi:2022fri,Karagiannis:2022ylq}. In this paper, we reassess this assumption and show that by not including non-Gaussian covariance in forecasts of the constraining power of the hydrogen bispectrum observed by a PUMA-like 21-cm intensity mapping experiment \cite{CosmicVisions21cm:2018rfq,Karagiannis:2019jjx}, one can underestimate the uncertainty in the linear regime by up to a factor of 
$\sim 5$ and $\sim 2$ for the local and equilateral type non-Gaussianity, respectively.
\\
\\
\textbf{Conventions \& Notation} We denote spatial vectors as $\bk{i}$ and its magnitude as $|\bk{i}| = k_i$. Sums of momenta are written as $\bk{1..n} = \sum_1^n \bk{i}$ e.g. $\bk{1}+\bk{2} = \bk{12}$. Momentum integrals are compactly written as $\int \frac{d^3\bk{i}}{(2\pi)^3} = \int_{\bk{i}}$
and $\delta_D$ denotes the Dirac delta function.
In order to compare to simulations of the matter bispectrum, our cosmology equals the fiducial cosmology of the \textsc{Quijote} suite \cite{Villaescusa-Navarro:2019bje}, which closely resembles the 2018 Planck constraints \cite{Planck:2018vyg}. For the analysis of the PUMA survey, we use the 2015 Planck constraints \cite{Planck:2015fie} to match previous forecasts.

\section{Theoretical Framework and Setup}
In order to estimate the signal-to-noise for high-redshift surveys observables, we need to introduce a few concepts. We are ultimately interested in constraining the early universe through primordial non-Gaussianities, thus we start off by defining correlations of primordial fluctuations, i.e. our signal of interest. Next, we introduce density perturbations, whose correlations at different positions in the sky are the building blocks of what we actually observe in high- (and low-) redshift surveys. Their dynamics driven by gravity determine the noise we need to overcome.

\subsection{Initial Conditions}
Quantum fluctuations during the inflationary epoch cause the expansion to end at slightly different times in different places, giving rise to tiny scalar density fluctuations $\zeta$ that source linear perturbations in the matter density field. In this way, linear fluctuations of the density field trace the primordial initial conditions of the universe. Even a small non-Gaussianity in the distribution of primordial fluctuations serves as an important way to discriminate between different models of inflation. Furthermore, it allows one to directly probe the particle content and interactions of the inflationary epoch \cite{Arkani-Hamed:2015bza,Lee:2016vti}. Since such non-Gaussianities are constrained to be small by CMB observations \cite{Akrami:2018odb}, in this work we consider only the first non-Gaussian statistic, which is the bispectrum. Hence, we require only the first two statistical moments of the primordial density distributions. In Fourier space, these are the power spectrum $P_{\zeta}(k)$ and bispectrum $B_{\zeta}(k_1,k_2,k_3)$, defined as
\begin{eqnarray}
\langle \zeta_{\bk{1}} \zeta_{\bk{2}} \rangle &=& (2\pi)^3 \delta_D(\bk{12}) P_\zeta(k_1),\\
\label{eq:zeta}
\langle \zeta_{\bk{1}} \zeta_{\bk{2}} \zeta_{\bk{3}} \rangle &=& (2\pi)^3 \delta_D(\bk{123}) B_\zeta(k_1,k_2,k_3).
\end{eqnarray}
 Different inflationary mechanisms give rise to distinct sizes and shapes of bispectra. It is customary to classify these bispectra into three main templates, the so-called local, equilateral and orthogonal templates, whose expressions are given in the Appendix, Eqs. \eqref{eq:local}, \eqref{eq:equi} and \eqref{eq:ortho}. The local shape typically arises in models of multi-field inflation and peaks in squeezed triangle configurations $k_1 \ll k_2 \sim k_3$. The equilateral shape is typically generated by self-interactions of the inflaton field and peaks for equilateral configurations $k_1 = k_2 = k_3$. Finally, the orthogonal shape, along with the equilateral one, is a natural prediction of the Effective Field Theory (EFT) of (single field) inflation \cite{Cheung:2007st} and peaks for both equilateral and flattened configurations $k_1 = k_2 + k_3$.

\subsection{Matter field and correlators}
The primordial initial conditions serve as the seed for the distribution of matter in the universe. We can therefore study the initial conditions of the universe by studying fluctuations of the matter density field, $\rho$, defined as $\delta(t,\textbf{x}) = \rho(t,\textbf{x})/\bar{\rho}(t)-1$, with $\bar \rho$ the mean density in a volume. Similar to the primordial case, we define correlations of $\delta(\mathbf{x})$ in Fourier space as
\begin{align}
\label{eq:pdelta}
\langle \delta_{\bk{1}} \delta_{\bk{2}}  \rangle &= (2\pi)^3 \delta_D(\bk{12}) P_\delta(k_1); \\
\langle \delta_{\bk{1}}  \delta_{\bk{2}}  \delta_{\bk{3}}  \rangle &= (2\pi)^3 \delta_D(\bk{123}) B_\delta(\bk{1},\bk{2},\bk{3}),\label{eq:matterb}\\
\langle \delta_{\bk{1}}  \delta_{\bk{2}}  \delta_{\bk{3}}  \delta_{\bk{4}} \rangle &= (2\pi)^3 \delta_D(\bk{1234}) T_\delta(\bk{1},\bk{2},\bk{3}, \bk{4}),\label{eq:mattert}
\end{align}
where we assume all fluctuations to be at equal times. In this work, we also need the $4$-point correlation function in Fourier space, known as the \emph{trispectrum}, for the computation of the non-Gaussian covariance. Even in the absence of a primordial bispectrum, or higher-order primordial correlators, fluctuations in the matter field grow via gravitational instability and become non-linear, thereby sourcing the matter bispectrum, trispectrum and higher-order correlations. The dynamical equations for $\delta$ describing this process can be solved perturbatively (see e.g.~\cite{Bernardeau:2001qr} for a review). This allows one to compute correlators analytically up to a mildy non-linear scale $k_{\rm NL}$. One way to estimate this scale is by computing
\begin{eqnarray}
\label{eq:knl}
k_\text{NL}(z) = \left[ \frac{1}{6\pi^2} \int_0^\infty dk \; P_\delta^L(k,z)\right]^{-1/2},
\end{eqnarray}
where $P_\delta^L$ is the linear matter power spectrum as defined in Eq.~\eqref{eq:matterp}. We use this scale to confine ourselves to the linear regime \footnote{Other definitions have been considered in the literature, e.g. \cite{Tomlinson:2022xud} studies the non-linear scale for the bispectrum specifically. The precise definition of $k_{\rm NL}$ does not qualitatively change the results of this paper.}. The induced bi- and trispectrum in this framework are presented in Eq. \eqref{eq:SPT-B} and Eq. \eqref{eq:SPT-T} of the Appendix. \\
To complement the perturbative approach, we resort to N-body simulations of the universe at large-scales solving the dynamical equations for $\delta$ numerically (see \cite{Angulo:2021kes} for a review). The advantage of N-body simulations is that they allow to directly measure correlations of $\delta$ even at non-linear scales, and to test analytic predictions. The drawback is that they are computationally expensive to run. We make use of publicly available \textsc{Quijote} simulations \cite{Villaescusa-Navarro:2019bje} for our estimates of signal-to-noise at low redshift (i.e. up to $z=3$). For higher redshifts, as the non-linear scale is pushed to very small scales, instead of fully solving dynamical equations we resort to \texttt{Monofonic} \cite{Michaux:2020yis}, which computes particle positions by solving third-order Lagrangian perturbation theory (3LPT) equations. Further details on how simulation data is used can be found in the Appendix.

\subsection{Fisher information and estimated uncertainty}
In this section, we introduce the quantities we use to estimate the uncertainty on the amplitude of primordial non-Gaussianity, $\fnl{}$, from observations of the bispectrum.
\paragraph{Fisher matrix.} A common way to quantify the information content of an observable is through the Fisher matrix. It encodes both the amount of information available from a measurement to constrain a parameter, as well as the correlation between different parameters. Given $N$ measurements of an observable, which for us will be the matter or hydrogen bispectrum, and a set of parameters we want to constrain, $\mathbf{p}$, the Fisher matrix is defined as
\begin{eqnarray}
\label{eq:fish}
F_{ab} = \sum_{TT'}\,\frac{\partial B_T}{\partial p_a}\left(C\right)_{TT'}^{-1}\frac{\partial B_{T'}}{\partial p_b},
\end{eqnarray}
where $T$ are triangle configurations in which the bispectra are measured, or calculated, $\mathbf{B}$ is the data vector of bispectra and $C_{TT'}$ is the covariance of $\mathbf{B}$, defined as
\begin{eqnarray}\label{eq:defcov}
C_{TT'} = \left\langle B_T\,B_{T'}\rangle - \langle B_T \rangle \langle B_{T'}\right \rangle.
\end{eqnarray}
The estimated uncertainty on a parameter $p_a$ is then defined as
\begin{eqnarray}
\sigma_{p_a} = \left(F^{-1}\right)^{1/2}_{a\,a},
\end{eqnarray}
where $F^{-1}$ indicates the matrix inverse of $F$.

\paragraph{N-body measurements.} When using numerical simulations, we measure the matter bispectrum on a finite size box with periodic boundary conditions, such that in this case $\delta_{\bk{}}$ is a discrete Fourier transform of the density contrast. The bispectrum estimator then is defined as \cite{Scoccimarro:1997st}
\begin{equation}\label{eq:estB}
\hat{B}(k_1,k_2,k_3)  \equiv  \frac{k_F^3}{N_{tr}}\sum_{\qv \in k}\,\delta_K(\qv_{123})\, \,\delta_{\qv_1}\,\delta_{\qv_2}\,\delta_{\qv_3},
\end{equation}
being $k_F=2\pi/L$ the fundamental frequency in a cubic box of side $L$, $N_{tr}$
gives the number of “fundamental triangles” formed by the
vectors $\qv_i$ satisfying the condition $\qv_{123} = 0$ that belong in the “triangle bin” defined by the triplet of
bin centers ($k_1, k_2, k_3$) and bin width $\Delta k$ \footnote{We also measure the power spectrum, since, as we show below, it enters in the calculation of the covariance. The estimator of the power spectrum is
\begin{equation}\label{eq:estp}
    \hat P(k) \equiv \frac{k^3_F}{N_k}\sum_{\mathbf{q} \in k} \delta_{\mathbf{q}}\,\delta_{-\mathbf{q}},
\end{equation}
where $N_k$ gives the number of modes in each k-bin.}.

The advantage of using N-body simulations is that the full  covariance can be estimated numerically from a sample of simulations using Eq.~\eqref{eq:defcov}, where now the average $\langle\cdot\rangle$ is over different realisations of the same simulation. 

It is also straightforward to compute the Gaussian contribution only, i.e. the case where different modes are uncorrelated. This contribution is given by the product of three power spectra \footnote{The approximate equality indicates the thin shell approximation.}
\begin{eqnarray}
\label{eq:ppp}
C^G_{TT'} \simeq \frac{(2\pi)^3\, k_F^3}{V_{123}}\,s_{123}\,\hat P(k_1)\hat P(k_2)\hat P(k_3)\, \delta_{TT'}
\end{eqnarray}
where $T,T'$ denote triangle bins, $V_{123} \simeq 8\pi^2 k_1 k_2 k_3 \Delta k^3$ is the volume of the bin,  $s_{123}=1,2,6$ for scalene, isosceles and equilateral triangles respectively and $\hat P(k_i)$ are power spectrum measurements. 
\paragraph{Limit of infinitely thin bins.} At high redshifts, the non-linear scale $k_{\rm NL}$ is pushed to smaller scales. At fixed bin width $\Delta k$, this implies a wider range of scales explored, and consequently a larger data vector and covariance. In order to keep the calculations within reasonable computational cost, one solution is to widen the range of bins, and to sample wavenumbers in log space. Alternatively, we choose to go in the limit of infinitely thin bins and promote the sums to integrals, such that the Fisher matrix becomes
\begin{equation}\label{eq:confish}
    F_{ab} = \int_{TT'} \frac{\partial B_T}{\partial p_a} C^{-1}_{TT'} \frac{\partial B_T'}{\partial p_b},
\end{equation}
where now the matter, or hydrogen, bispectrum are estimated using perturbation theory, as explained in the Appendix.
Calculating Eq.~\eqref{eq:confish} now implies knowledge of the dependence on triangle configurations $T,T'$ of the inverted full covariance matrix, which is typically hard to compute. In Appendix (Eqs.~\eqref{eq:neumann} to \eqref{eq:fabappr}) we outline a strategy that is based on splitting the covariance into Gaussian and non-Gaussian contributions, $C = C_{\rm G} + C_{\rm nG}$, and expanding the inverse as a Neumann series. We then approximate this series such that the Fisher matrix in the limit of thin bins becomes:
\begin{eqnarray}\label{eq:expfish}
F_{ab}  =  \frac{\left(F_{ab}^{\rm G}\right)^2}{F_{ab}^{\rm G} + \delta F_{ab}^{\rm nG}},
\end{eqnarray}
where $F_{ab}^{\rm G}$ is the Fisher matrix computed using only the Gaussian covariance $C_{\rm G}$ to compute the inverse covariance and $\delta F_{ab}^{\rm nG}$ is the non-Gaussian correction computed using as inverse the product of matrices $-C_{\rm G}^{-1} C_{\rm nG} C^{-1}_{\rm G}$.
\paragraph{Model for the non-Gaussian covariance.}
The goal of this paper is to compute how $\sigma$ varies for $f_{\rm NL}$ whether we are considering only the Gaussian term $C_{\rm G}$ or a more complete modelling of the covariance including non-Gaussian terms. As explained above, when using N-body simulations, the Gaussian and the full covariance are computed numerically. In the case of thin bins, we need to introduce a model of the bispectrum covariance. Inserting Eq.~\eqref{eq:estB} into Eq.~\eqref{eq:defcov}, the computation involves the correlator of $6$ fields in Fourier space, which can be combined in four different ways: the Gaussian term is the product of three power spectra (`PPP' term), given by Eq.~\eqref{eq:ppp}. Non-Gaussian terms are represented by either the product of two bispectra (`BB' term), or the product of a power spectrum and a trispectrum (`PT' term), or finally the connected $6$-point function, the so-called \emph{pentaspectrum}. The pentaspectrum is negligible in most practical cases (see \cite{Biagetti:2021tua} for a rough estimate). The key point of this paper is to account for the `BB' and `PT' terms in signal-to-noise estimates at high redshifts. The `BB' term, again assuming that correlators are slowly varying in the $k$-shells, can be written as
 \begin{equation}
     C_{\rm nG}^{\rm BB} \simeq B_T B_{T'} (\Sigma_{TT'}^{11} + 8\,\,{\rm perm}),
\end{equation}
where $\Sigma^{ij}_{TT'}$ is a mode-counting factor that again depends on the shape of the triangles. The `PT' term is similarly written. We calculate these terms for the matter bispectrum predictions using Eqs. \eqref{eq:matterp}, \eqref{eq:SPT-B} and \eqref{eq:SPT-T}. For the hydrogen bispectrum we use the following model for the covariance
\begin{equation}\label{eq:modelcov}
     C \approx C_{\rm G}+ 2\,C_{\rm nG}^{\rm BB},
 \end{equation}
 where the `PT' term is approximated to be equal to the `BB' term, which is a good approximation for squeezed triangles for which the non-Gaussian terms are largest \cite{Biagetti:2021tua}. 

\section{Constraining $\fnl{}$ at high redshifts}
The primary goal of this work is to show the importance of including non-Gaussian terms in the covariance when estimating the uncertainty to the primordial non-Gaussian amplitude $\fnl{}$ in high-redshift surveys. 

One could be tempted to neglect the non-Gaussian covariance at high redshifts on scales larger than $k_{\rm NL}$ at that redshift. In this linear regime one might expect modes of different wavelength to be mostly uncorrelated, such that the covariance is diagonal and Gaussian terms dominate. As we show in what follows, this intuition fails: off-diagonal terms become important well within what is typically considered the linear regime based on Eq.~\eqref{eq:knl}.

\subsection{Uncertainty on $\fnl{}$ from the matter bispectrum}\label{sec:matter}

As a testing ground, we first consider the matter bispectrum in real space as our observable and compute the estimated uncertainty of the primordial non-Gaussian amplitude $\fnl{}$ for the primordial bispectra of the local, equilateral and orthogonal type as defined previously. In this setup, $\fnl{}$ is the only parameter. When using finite-sized bins, we evaluate the derivative $\partial \mathbf{B} / \partial f_{\rm NL}$ averaging over the bins (see Eq. \eqref{eq:binavgB} of the Appendix), while in the case of infinitely thin bins the derivative is analytically computed directly from the templates Eqs. \eqref{eq:local}, \eqref{eq:equi} and \eqref{eq:ortho}.

\begin{figure}
    \includegraphics[scale=0.5]{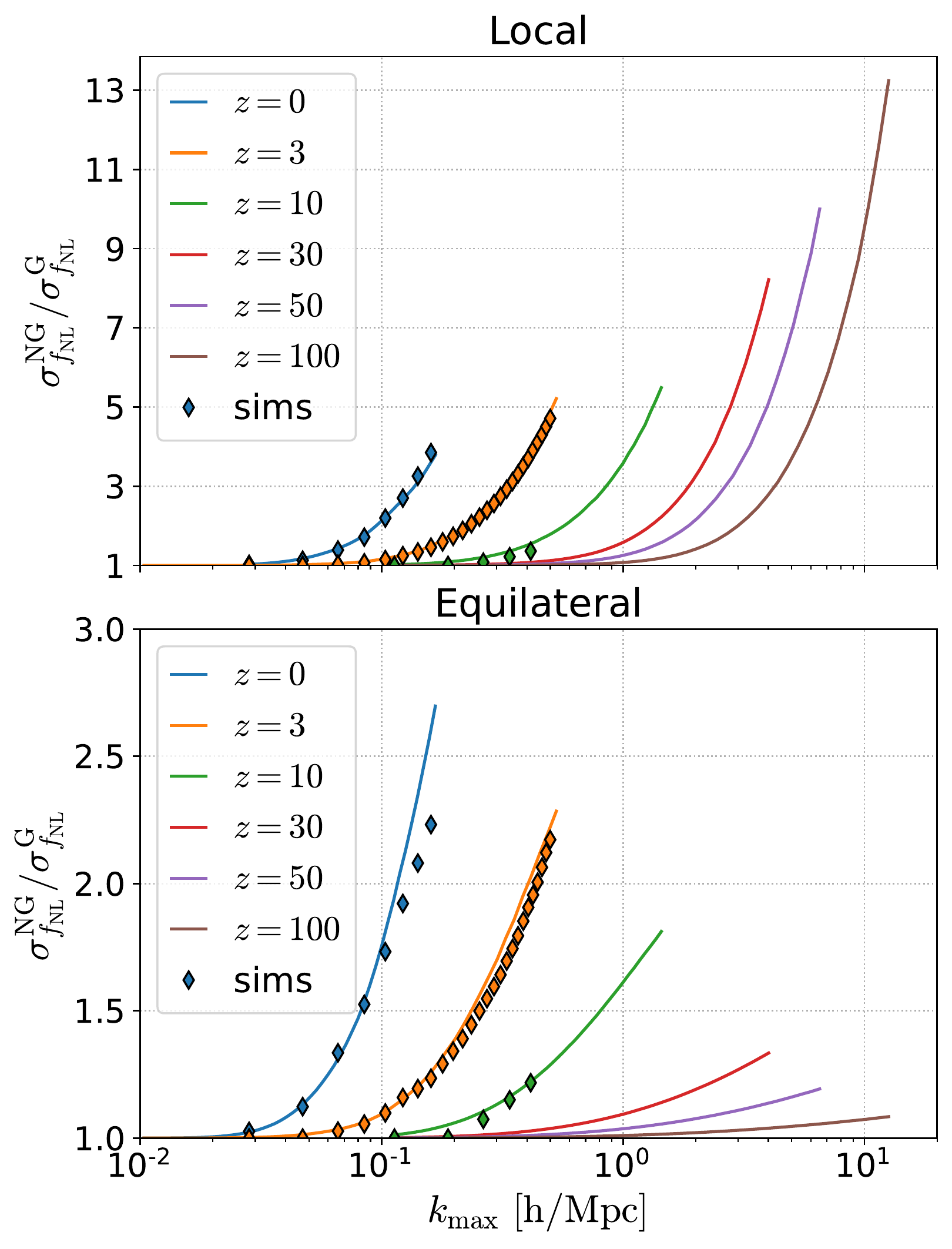}
    \caption{Estimate of relative increase in error on $f_{\rm NL}$ due to non-Gaussian covariance as a function of $\kmax{}$. The diamonds present results obtained using the \textsc{Quijote} simulations ($z=0,3$) or 3LPT ($z = 10$). Note the different scales on the vertical axes. The local bispectrum is expected to be significantly affected when accounting for non-diagonal covariance even at very high redshifts. Solid lines are estimated up to the non-linear scale $k_{\rm NL}$ at each redshift. For $z=0$ and $3$ the simulation results are also shown up to the non-linear scale, while for $z=10$ they are shown up to the scale at which shot-noise becomes a significant contribution to the covariance.}
    \label{fig:R_BS}
\end{figure}

Figure \ref{fig:R_BS} shows the ratio of the estimated uncertainty computed with non-Gaussian over Gaussian covariance as a function of the maximum wavenumber $k_{\rm max}$. The uncertainty computed using the infinitely thin bin approximation is shown in solid lines, while simulation measurements are shown as diamonds. Solid lines are computed up to the non-linear scale $k_{\rm NL}$ at that redshift. The uncertainty on local type non-Gaussianity is most affected by the introduction of off-diagonal covariance, increasing by a factor of $\sim5$ at $k_{\rm max} \approx k_{\rm NL}$ for redshifts lower than $z=10$ and even higher at higher redshifts. This is because the off-diagonal covariance is largest for squeezed triangle configurations where the local type non-Gaussianity has most of its signal \cite{Biagetti:2021tua} \footnote{This is very similar to how lensing-induced covariance mostly affects measurements of the local bispectrum in the CMB \cite{Coulton:2019odk}}. Equilateral non-Gaussianity is less affected, since most of its signal comes from equilateral triangle configurations whose covariance is large only when approaching non-linear scales. Still, the loss is almost a factor of $2$ at $z\lesssim 10$. For a discussion and the results of the orthogonal bispectrum we refer to the Appendix.

It is important to note that these results do not imply that the uncertainty does not improve overall, since we are still able to access more modes as we increase $\kmax{}$. Rather, our analysis shows that off-diagonal non-Gaussian covariance reduces the amount of information gained by probing smaller scales, or in other words, the signal-to-noise saturates. We show a clear representation of this fact in Figure \ref{fig:snbs} in the Appendix, where we directly plot the uncertainty of $\fnl{}$ including non-Gaussian terms as a function of $k_{\rm max}$ at different redshifts for a fictitious matter survey.

\subsection{Uncertainty on $\fnl{}$ from the hydrogen bispectrum}\label{sec:hydro}
To make contact with actual future observations, we consider a realistic PUMA-like intensity mapping survey setup. 
PUMA is a proposed 21-cm intensity mapping experiment aimed at measuring the distribution of neutral hydrogen through the 21-cm hyperfine transition between redshift 2 and 6. One of the key science drivers for PUMA is to provide better constraints on primordial non-Gaussianity with respect to the CMB \cite{CosmicVisions21cm:2018rfq} (see also Figure 5 in \cite{Achucarro:2022qrl} for a comparison to other future surveys). \\
As compared to the simplified scenario considered in Figure \ref{fig:R_BS}, the calculation of the estimated uncertainty in this case involves several complications. First of all, neutral hydrogen is a biased tracer of the matter field. This introduces additional non-linearities and we need to define a set of nuisance (bias) parameters that are fixed through observations (see \cite{Desjacques:2016bnm} for a review). Secondly, we need to compute correlators in redshift space, taking into account the survey geometry and foregrounds. Lastly, the presence of primordial non-Gaussianity introduces additional bias parameters. This last effect famously appears already at the power spectrum level for the local template, known as scale dependent bias (\cite{Dalal:2007cu, Matarrese:2008nc, Slosar:2008hx} and \cite{Biagetti:2019bnp} for a recent review). For this reasons, forecasts of $\sigma(f_{\rm NL})$ depend sensitively on many assumptions, and would need to include the tracer power spectrum in order to be realistic. Here we limit ourselves to calculate the uncertainty using the tracer bispectrum only, rather than performing a full forecast, since our goal is to show the loss of constraining power due to the inclusion of non-Gaussian covariance on the bispectrum \footnote{We have confirmed that our forecasts, using very similar assumption about the PUMA survey, result in forecasts on $\sigma(f_{\rm NL})$ that are consistent with those presented in Refs.~\cite{Karagiannis:2019jjx,Sailer:2021yzm} when neglecting non-Gaussian covariance}. For our computations, we follow the setup presented in \cite{Karagiannis:2019jjx}. We take into account foreground noise, which effectively limits the largest scale accessible by the survey in each redshift bin, as described in the Appendix Eqs.~\eqref{eq:kpar} and \eqref{eq:kper}. In this setup, beside $f_{\rm NL}$ the hydrogen bispectrum is a function of $7$ parameters: three bias parameters, two shot-noise parameters, the dimensionless linear growth rate $f$ and the velocity dispersion $\sigma_v$. We compute the fiducial value of these parameters as a function of redshift following \cite{Castorina:2016bfm,Karagiannis:2019jjx} and the expression for the hydrogen bispectrum is found in the Appendix, Eq.~\eqref{eq:hibis}. We also calculate the hydrogen power spectrum, given in Eq.~\eqref{eq:hipow}, as we use it to compute the Gaussian covariance. To compute the non-Gaussian covariance, we use the model of Eq. \eqref{eq:modelcov}.
We then proceed in computing the Fisher matrix, which is estimated in the thin bins form of Eq. \eqref{eq:expfish}. 
We marginalise over all the $7$ nuisance parameters entering the bispectrum as described in Eq. \eqref{eq:marginal} of the Appendix. 

Figure \ref{fig:PUMA} shows the ratio of the estimated uncertainty computed using a non-Gaussian covariance over a Gaussian approximation for the local and equilateral type non-Gaussianities as a function of redshift for a PUMA-like experiment. We compute the uncertainty for two different values of $k_{\rm max}$, corresponding to $0.5\, k_{\rm NL}$ (dashed lines) and $0.75\, k_{\rm NL}$ (solid lines), as we expect Eq. \eqref{eq:knl} to be less accurate for tracers. Our results show that even for a more conservative choice of $k_{\rm max} = 0.5\, k_{\rm NL}$, the effect is significant. The increase in uncertainty ranges from a factor of $2$ to a factor of $5$ for local type non-Gaussianity. We therefore conclude previous forecasts on constraining $\fnl{}$ at high-redshift are too optimistic \cite{Munoz:2015eqa,Meerburg:2016zdz,Karagiannis:2019jjx,Floss:2022grj,Yamauchi:2022fri} and non-Gaussian covariance will have to be considered in order to produce more realistic forecasts. A similar estimation for a generic biased tracer was performed in  \cite{dePutter:2018jqk} up to $z=10$ and shows qualitative agreement with Figure \ref{fig:PUMA}.

\begin{figure}
    \includegraphics[scale=0.5]{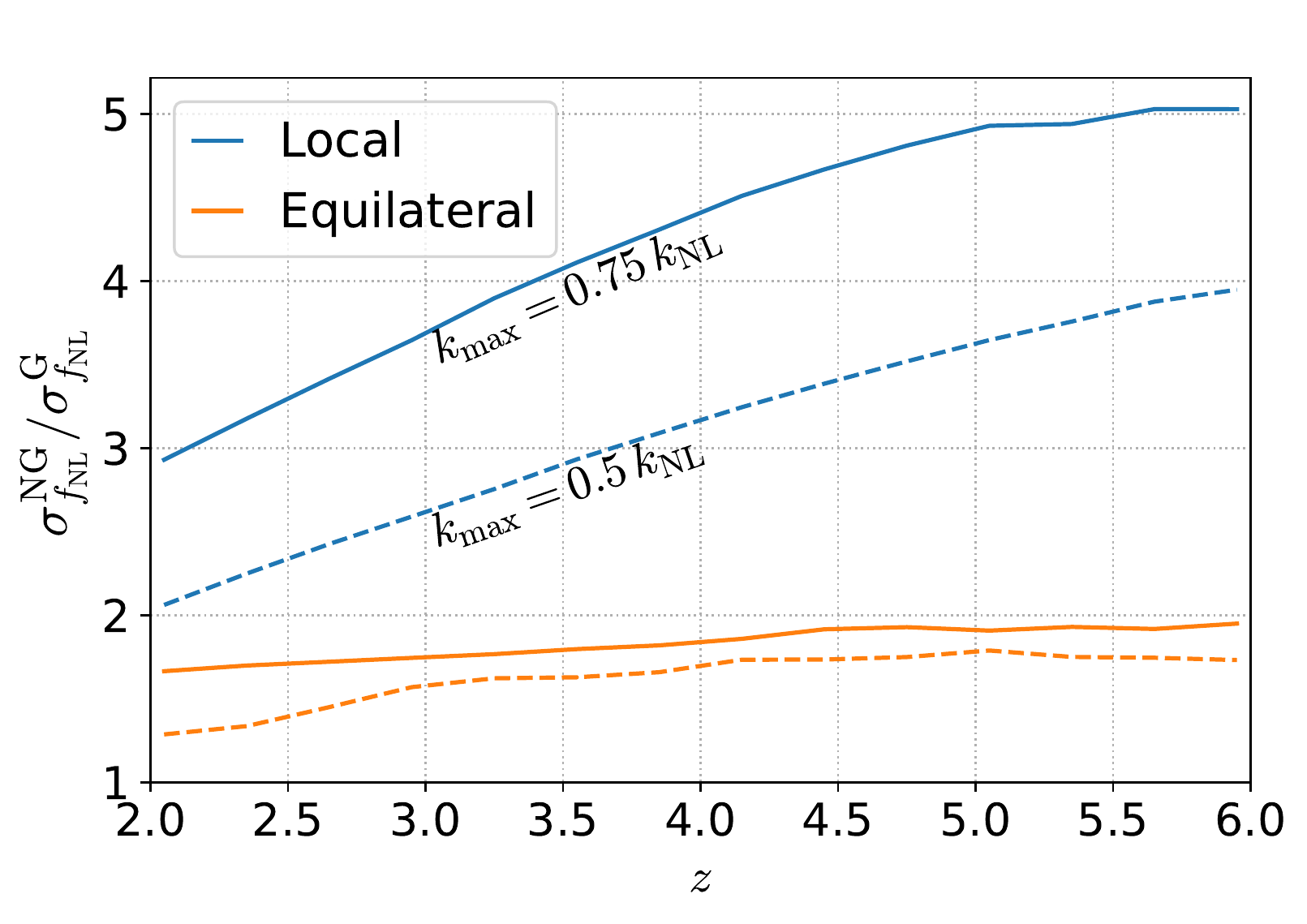}
     \caption{Estimate of relative increase in error on the non-Gaussian amplitude $f_{\rm NL}$ due to non-Gaussian covariance of the hydrogen bispectrum, as a function of redshift for a PUMA-like experiment when marginalising over the $7$ additional parameters of the hydrogen bispectrum. We show the results for $\kmax{} = 0.75\, k_{\rm NL}$ (solid lines) and $\kmax{} = 0.5\, k_{\rm NL}$ (dashed lines). 
    }
    \label{fig:PUMA}
\end{figure}

\section{Discussion and Conclusions}
We studied the impact of non-Gaussian terms in the covariance on  measurements of cosmological correlators. Specifically, we aim to quantify the effect on the estimated uncertainty of the primordial non-Gaussian amplitude $\fnl{}$ when using the bispectrum at high redshift as an observable. Because off-diagonal components are small compared to the diagonal, most studies have typically neglected this covariance. We showed that, when looking at the information content, there is a significant impact on the constraining power on primordial non-Gaussianity due to this non-Gaussian mode coupling, even at high redshifts and well below the non-linear scale as defined in Eq.~\eqref{eq:knl}. 

We have first computed the effect of non-Gaussian covariance on $\sigma_{\fnl{}}$ using the matter bispectrum in real space and then performed a more realistic estimation using the hydrogen bispectrum as measured from a PUMA-like experiment. This proposed 21-cm intensity mapping experiment has the potential to constrain primordial bispectra to reach beyond constraints set by the CMB. Yet, our analysis shows that not accounting for the full covariance can overestimate the constraining power of the hydrogen bispectrum measured by PUMA up to a factor of $5$ for local type non-Gaussianity and $2$ for equilateral. For local type non-Gaussianity, the primary observable is actually the tracer power spectrum, thanks to the so-called scale dependent bias, which we do not include in our analysis. Nevertheless, our results imply that combining it with the bispectrum does not help as much as it is expected to considering a Gaussian covariance only. Moreover, they motivate including the bispectrum-power spectrum cross covariance in the joint analysis, which is also a non-Gaussian contribution \cite{Biagetti:2021tua}.
Overall, our result suggests we should reconsider some of the existing forecasts and make sure the projected numbers are not overly optimistic for future high redshift surveys such as PUMA, MegaMapper \cite{Schlegel:2019eqc} and the Maunakea Spectroscopic Explorer \cite{MSEScienceTeam:2019bva}. 

Future constraints on primordial non-Gaussianities will depend on our ability to extract information from large scale structures. Intuitively, the main obstacle to constrain primordial spectra is set by the non-linear scale which estimates when loop-corrections become important. Here we show that for the Fisher information on $\fnl{}$ it is important to account for non-Gaussian bispectrum covariance, even for modes that are still considered linear. The results are comparable to what was found for measurements of the CMB bispectrum, where lensing induced off-diagonal covariance is the main limitation as we start to measure smaller scales and increase the number of accessible modes.

For the CMB, it was shown that the lensing induced covariance can be accounted for by delensing the data before applying the standard estimators \cite{Coulton:2019odk}. The analogy here would be to ``degravitate'' the data, a technique that is well established in studies of the Baryon Acoustic Oscillations in galaxy surveys \cite{Eisenstein:2006nk}. It might be possible to explore this option at high redshifts, where the physics is still perturbatively tractable. At lower redshifts however, it likely suggests that existing estimators are sub-optimal or that we have adopted inefficient summary statistics that need to be revisited. Similar conclusions were drawn in Ref.~\cite{Coulton:2022qbc}. Applying reconstruction methods \cite{Leclercq:2014fta} or using simulations (e.g. through simulation based inference \cite{cranmer2020frontier}) \cite{Alsing:2018eau,Alsing:2019xrx,Jeffrey:2020itg,Miller:2020hua,Cole:2021gwr}, both active fields of investigation, will certainly help to establish to what degree we have to modify our analysis tools in search for signs of primordial non-Gaussianity.  

The code used to produce the results in this work is publicly available \footnote{\url{https://github.com/tsfloss/pyNG}}.

\section*{Acknowledgements}
The authors would like to thank Emanuele Castorina, Will Coulton, Simon Foreman, Dionysios Karagiannis, Emiliano Sefusatti and Anže Slosar for useful discussions and comments on a draft. While finalising this work, we became aware of related work by Coulton et al. \cite{Coulton:2022qbc} and thank the authors for sharing their draft ahead of submission. We would also like to thank Francisco Villaescusa-Navarro and the whole \textsc{Quijote} team for making the simulation suite available. We thank the Center for Information Technology of the University of Groningen for their support and for providing access to the Peregrine high performance computing cluster. T.F is supported by the Fundamentals of the Universe research program within the University of Groningen. M.B acknowledges support from the Netherlands Organization for Scientific Research (NWO), which is funded by the Dutch Ministry of Education, Culture and Science (OCW) under VENI grant 016.Veni.192.210.  P.D.M acknowledges support from the Netherlands organization for scientific research (NWO) VIDI grant (dossier 639.042.730).

\appendix

\section{Primordial Bispectra}\label{app:pb}
The local, equilateral and orthogonal bispectrum templates are given by:
\begin{eqnarray}
\label{eq:local}
B^\text{loc}_\zeta(k_1,k_2,k_3) &=& \frac{6}{5} f_{\text{NL}}^{\text{loc}}\left( P_{\zeta,1} P_{\zeta,2} + P_{\zeta,1} P_{\zeta,3} + P_{\zeta,2} P_{\zeta,3} \right) \nonumber \\ \\
\label{eq:equi}
B^\text{equil}_\zeta(k_1,k_2,k_3) &=& \frac{18}{5} f_{\text{NL}}^{\text{equil}}\Big(-(P_{\zeta,1} P_{\zeta,2} + \text{2 perms.}) - \nonumber \\
                            && 2 P_{\zeta,1}^{2/3} P_{\zeta,2}^{2/3} P_{\zeta,3}^{2/3} +  \nonumber \\ 
                            && \left. (P_{\zeta,1}^{1/3} P_{\zeta,2}^{2/3} P_{\zeta,3} + \text{5 perms.}) \right) \\
\label{eq:ortho}
B^\text{ortho}_\zeta(k_1,k_2,k_3) &=& \frac{18}{5} f_{\text{NL}}^{\text{ortho}}\Big(-3 (P_{\zeta,1} P_{\zeta,2} + \text{2 perms.}) - \nonumber \\
                            && 8 P_{\zeta,1}^{2/3} P_{\zeta,2}^{2/3} P_{\zeta,3}^{2/3} +  \nonumber \\ 
                            && \left. 3 (P_{\zeta,1}^{1/3} P_{\zeta,2}^{2/3} P_{\zeta,3} + \text{5 perms.}) \right),
\end{eqnarray}
where we introduced the shorthand notation $P_{\zeta,1} = P_\zeta(k_1)$. Note that although Eq.~\eqref{eq:zeta} demonstrates that the primordial bispectrum can in principle depend on the full three-momenta, for the above primordial shapes there is no angular dependence and they only depend on the magnitudes of the triangle's momenta (i.e. the shape of the triangle).

\section{Standard Perturbation Theory at Tree Level}
Within the perturbative regime, gravitational interactions can be treated within the framework of Standard Perturbation Theory (SPT). For an extensive review we refer to e.g. \cite{Bernardeau:2001qr}. Here we will present only the results relevant to our work. Since we require the gravitational trispectrum, we expand the density field to third order:
\begin{eqnarray}
\delta_{\bk{}}(z) = \delta^{(1)}_{\bk{}}(z) + \delta^{(2)}_{\bk{}}(z) + \delta^{(3)}_{\bk{}},
\end{eqnarray}
where the super-script denotes the order of the perturbation. Solving the evolution equations order by order in perturbations one finds (dropping the explicit time-dependence):
\begin{eqnarray}
\delta^{(2)}_{\bk{}} &=& \int_{\bq{}} F_2(\bq{},\bk{}-\bq{})\delta^{(1)}_{\bq{}}\delta^{(1)}_{\bk{}-\bq{}}, \nonumber \\
\delta^{(3)}_{\bk{}} &=& \int_{\bq{1},\bq{2}} F_3(\bq{1},\bq{2},\bk{}-\bq{12})  \delta^{(1)}_{\bq{1}}\delta^{(1)}_{\bq{2}}\delta^{(1)}_{\bk{}-\bq{12}}.
\end{eqnarray}
These higher order perturbations will induce the gravitational correlations. At tree level then, the bispectrum of the density field as due to gravitational interactions, is found to be:
\begin{eqnarray}
\label{eq:SPT-B}
B^{\rm snG}_\delta(\bk{1},\bk{2},\bk{3}) &=& \langle \delta^{(1)}_{\bk{1}} \delta^{(1)}_{\bk{2}} \delta^{(2)}_{\bk{3}} \rangle + \textrm{ 2 perms.} \nonumber \\ &=& 2F_2(\bk{1},\bk{2})P^L_\delta(k_1)P^L_\delta(k_2) + \textrm{2 perms.}, \nonumber \\
\end{eqnarray}
where the linear power spectrum is defined as:
\begin{eqnarray}\label{eq:matterp}
\langle \delta^{(1)}_{\bk{1}} \delta^{(1)}_{\bk{2}} \rangle &=& (2\pi)^3 \delta_D(\bk{12})P_\delta^L(k_1) \nonumber \\
&=& (2\pi)^3 \delta_D(\bk{12})\mathcal{M}(k_1,z)P_\zeta(k_1).
\end{eqnarray}
Here $\mathcal{M}$ is the linear transfer function. The tree level trispectrum consists of two contributions:
\begin{eqnarray}
\label{eq:SPT-T}
T^{\rm snG}_\delta(\bk{1},\bk{2},\bk{3},\bk{4}) &=& T^{1122}_\delta(\bk{1},\bk{2},\bk{3},\bk{4}) \nonumber \\ &&+ T^{1113}_\delta(\bk{1},\bk{2},\bk{3},\bk{4})
\end{eqnarray}
with the two contributions given by:
\begin{eqnarray}
T^{1113}_\delta(\bk{1},\bk{2},\bk{3},\bk{4}) &=& \langle \delta^{(1)}_{\bk{1}}\delta^{(1)}_{\bk{2}}\delta^{(1)}_{\bk{3}}\delta^{(3)}_{\bk{4}}\rangle + \textrm{ 3 perms.} \nonumber \\ &=&  6F_3(\bk{1},\bk{2},\bk{3})\prod_{i=1}^3 P^L_\delta(k_i)\nonumber \\
&&+ \textrm{ 3 perms.} \nonumber \\ 
T^{1122}_\delta(\bk{1},\bk{2},\bk{3},\bk{4}) &=& \langle \delta^{(1)}_{\bk{1}}\delta^{(1)}_{\bk{2}}\delta^{(2)}_{\bk{3}}\delta^{(2)}_{\bk{4}}\rangle + \textrm{ 5 perms.} \nonumber \\
&+& 4\left[F_2(-\bk{1},\bk{13})P^L_\delta(k_{13}) \right. \nonumber \\ &+& \left. F_2(-\bk{1},\bk{14})P^L_\delta(k_{14})\right]\nonumber \\\ && \times P^L_\delta(k_1)P^L_\delta(k_2) \nonumber \\ &+& \textrm{5 perms.}
\end{eqnarray}
The kernels $F_2,F_3$ determine how modes of different wavelength are coupled by gravity. For their explicit form, see e.g. \cite{Bernardeau:2001qr}.
Beyond tree level, the higher order perturbations will also induce a correction to the power spectrum, known as the 1-loop power spectrum:
\begin{eqnarray}
\langle \delta^{(2)}_{\bk{1}}\delta^{(2)}_{\bk{2}}\rangle = (2\pi)^3 \delta_D(\bk{12})P_\delta^{\textrm{1-loop}}(k_1).
\end{eqnarray}

\section{Details on simulations}
In order to verify our computations of non-Gaussian covariance, we compare our results to simulations. For redshifts $z=0$ and $3$ we use the \textsc{Quijote} simulation suite \cite{Villaescusa-Navarro:2019bje}. \textsc{Quijote} consists of 15000 N-body simulations using a fiducial cosmology, enough to obtain an accurate covariance matrix up to the scales of interest in this work. The simulations consist of $512^3$ particles in a box with sides $1000$  Mpc/h, setting the fundamental mode to $k_F = (2\pi/1000) \approx 0.0063$ h/Mpc. To construct the covariance matrix, we use the power spectrum and bispectrum measurements as provided in the suite, which use a binning of $\Delta k = 3\, k_F$, $\kmin{} = \frac{3}{2}k_F$ and an interpolation grid of size $360^3$. For redshift $z=10$ we use $\sim 4000$ realisations of initial conditions (ICs) generated using third order Lagrangian perturbation theory (3LPT) with the \texttt{Monofonic} code \cite{Michaux:2020yis}. Since at higher redshifts the power spectrum and bispectrum are smaller, shot-noise becomes increasingly dominant. The $z=10$ realisations are therefore generated with $512^3$ particles in a box of $250$ Mpc/h, setting the fundamental mode to $k_F = (2\pi/250) \approx 0.025$ h/Mpc. Measurements of the power spectrum and bispectrum are done using the codes \texttt{Pylians} \footnote{\url{https://github.com/franciscovillaescusa/Pylians}} and \texttt{PySpectrum} \footnote{\url{https://github.com/changhoonhahn/pySpectrum}} respectively, with the same settings as used for the \textsc{Quijote} measurements. Having measured the power spectrum and bispectrum from many realizations of the simulation, we obtain the covariance matrix through Eq. \eqref{eq:defcov}. When inverting the covariance matrix from simulations, we include the Hartlap factor to unbias the numerical matrix \cite{Hartlap:2006kj}.

\paragraph{Evaluation of bin-averaged primordial bispectra.}

In order to compute the Fisher information for primordial bispectra from the simulations, we need the bin-averaged derivatives of the theoretical bispectrum with respect to $\fnl{}$ that appear in Eq. \eqref{eq:fish}:
\begin{align}
\label{eq:binavgB}
\frac{\partial \hat{B}(k_1,k_2,k_3)}{\partial \fnl{}} &= \frac{1}{V_{123}}\int_{k_1} d^3 \bq{1} \int_{k_2}d^3 \bq{2} \int_{k_3}d^3 \bq{3} \nonumber \\ & \frac{\partial B(\bq{1},\bq{2},\bq{3})}{\partial \fnl{}} \nonumber  \\ 
 &= \frac{1}{V_{123}}\int_{k_1} d^3 \bq{1} \int_{k_2}d^3 \bq{2} \int_{k_3}d^3 \bq{3} \nonumber \\ & \left(\prod_{i=1}^3\mathcal{M}(q_i,z)\right) B_\zeta(\bq{1},\bq{2},\bq{3})|_{\fnl{}=1},
\end{align}
where the hat denotes a bin-averaged quantity and the volume of the bin is given by
\begin{eqnarray}
V_{123} = \int_{k_1} d^3 \bq{1} \int_{k_2}d^3 \bq{2} \int_{k_3}d^3, \bq{3}
\end{eqnarray}
and the integrals denote a binning similar to that of the simulation measurements, i.e. over spherical shells with centers $k_i$ and width $[k_i-3k_F/2, k_i +3k_F/2]$. 

\begin{figure}
    \includegraphics[scale=0.50]{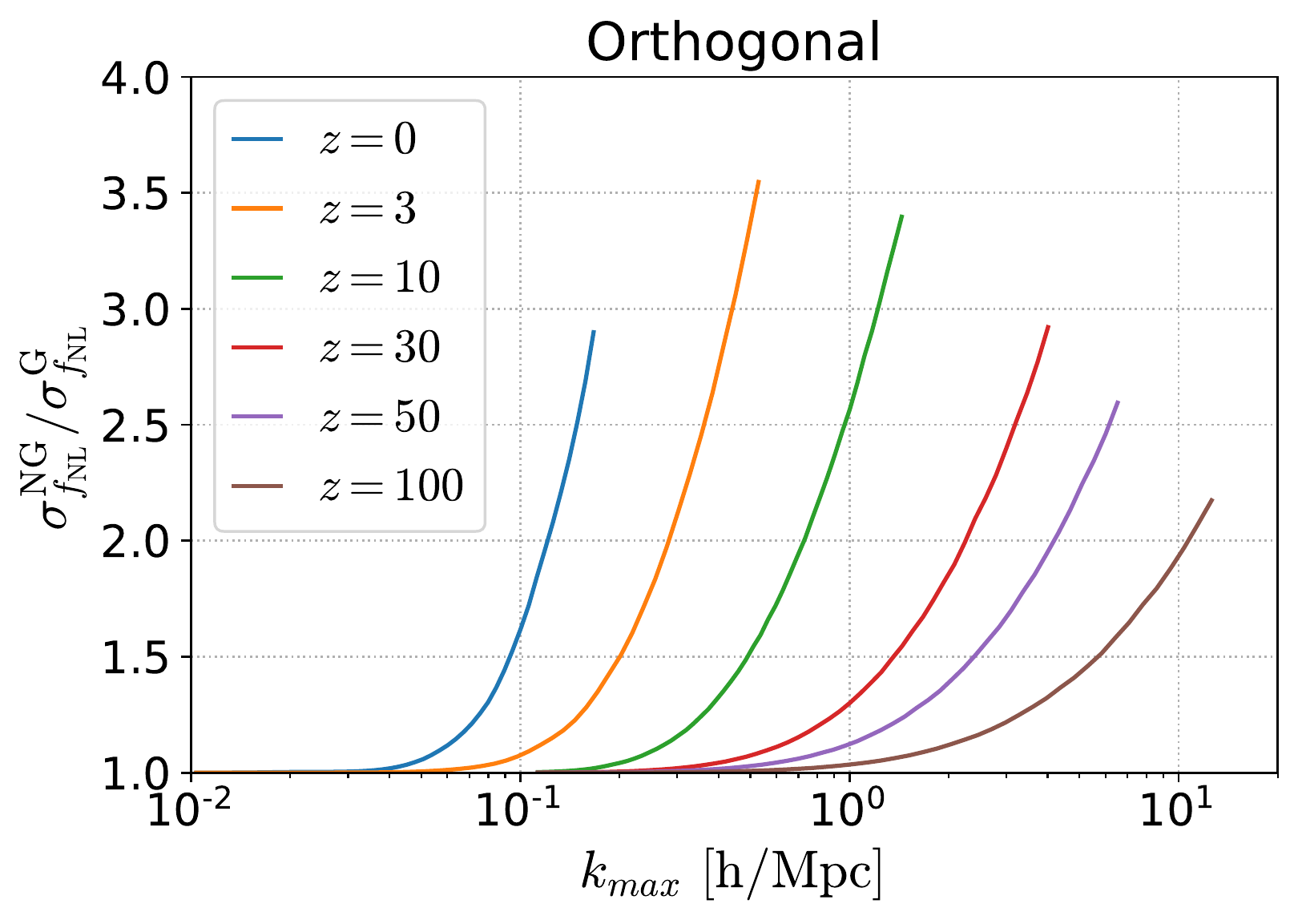}
    \caption{Similar plot as Figure \ref{fig:R_BS} but now for orthogonal non-Gaussianity.}
    \label{fig:R_Orth}
\end{figure}

\paragraph{Orthogonal shape.}
When computing the bin-averaged orthogonal shape bispectrum, one realizes that it becomes negative for certain triangle configurations. For the coarse binning ($\Delta k = 3k_F$) of the simulation data we use, the bin-averaged bispectrum in Eq.~ \eqref{eq:binavgB} suffers from cancellations within the bin, removing part of the signal. Since remeasuring the bispectrum in \textsc{Quijote} simulations with finer binning goes beyond the scope of this paper, we decided to omit these results. Nonetheless, our predictions in the thin bin approximation are presented in Figure \ref{fig:R_Orth} and for $z=0$ agree well with the simulation results presented in \cite{Coulton:2022qbc}.

\section{Details on the Fisher matrix in the thin bins limit}
To produce our theoretical predictions of the loss of constraining power due to non-Gaussian covariance, we take the continuous limit of Eq. \eqref{eq:fish} while approximating the inverse covariance matrix. 

As explained in the main text, this allows us to probe a wide range of scales at low computational cost.
The main complication for Eq. \eqref{eq:confish} is to explicitly compute the inverse covariance as a function of triangle configurations. To this end, we expand the inverse using a Neumann series:
\begin{eqnarray}
\label{eq:neumann}
C^{-1} &=& \sum_{n=0}^{\infty} \left(-C_{\rm G}^{-1} C_{\rm nG}\right)^n C_{\rm G}^{-1},
\end{eqnarray}
such that we never have to invert the non-Gaussian covariance matrix that contains off-diagonal terms. Hence the Fisher matrix becomes the sum of infinitely many terms:
\begin{equation}
F_{ab} = \sum_{n=0}^{\infty} \sum_{TT'}\frac{\partial B_T}{\partial p_a}  \left(\left(-C_{\rm G}^{-1} C_{\rm nG}\right)^n C_{\rm G}^{-1}\right)_{TT'} \frac{\partial B_T}{\partial p_b}.
\end{equation}
Since the terms in this sum are increasingly complicated to compute (in the continuous limit, every matrix multiplication becomes an integral over triangle configurations), we choose to approximate the expansion using:
\begin{eqnarray}\label{eq:fabappr}
F_{ab} &=& \sum_{TT'}\frac{\partial B_T}{\partial p_a} \left(C_{\rm G}^{-1}\right)_{TT'} \frac{\partial B_T}{\partial p_b} \nonumber \\ &&+ \sum_{n=1}^{\infty} \frac{\left(\sum_{TT'}\frac{\partial B_T}{\partial p_a}  \left(-C_{\rm G}^{-1} C_{\rm nG}C_{\rm G}^{-1}\right) _{TT'} \frac{\partial B_T}{\partial p_b} \right)^n}{\left(\sum_{TT'}\frac{\partial B_T}{\partial p_a} \left(C_{\rm G}^{-1}\right)_{TT'} \frac{\partial B_T}{\partial p_b}\right)^{n-1}}\nonumber \\ &=& F_{ab}^{G} + \sum_{n=1}^{\infty}(-1)^n\frac{\left(\delta F_{ab}^{\rm nG}\right)^{n}}{\left(F_{ab}^{\rm G}\right)^{n-1}}
\end{eqnarray}
which can be recognised as the expansion of Eq. \eqref{eq:expfish}.
The approximation of Eq. \eqref{eq:fabappr} seems to work reasonably well in the range of $k_{\rm max}$ we compare with simulation results (see Fig. \ref{fig:R_BS} in the main text). On the other hand, our key result is that these non-Gaussian terms are actually important, therefore we expect our approximation to break down. This motivates further work in defining a proper estimator for primordial non-Gaussianity in the presence of non-Gaussian covariance terms.

\section{Dependence of the uncertainty on $k_{\rm max}$}

\begin{figure}
    \includegraphics[scale=0.50]{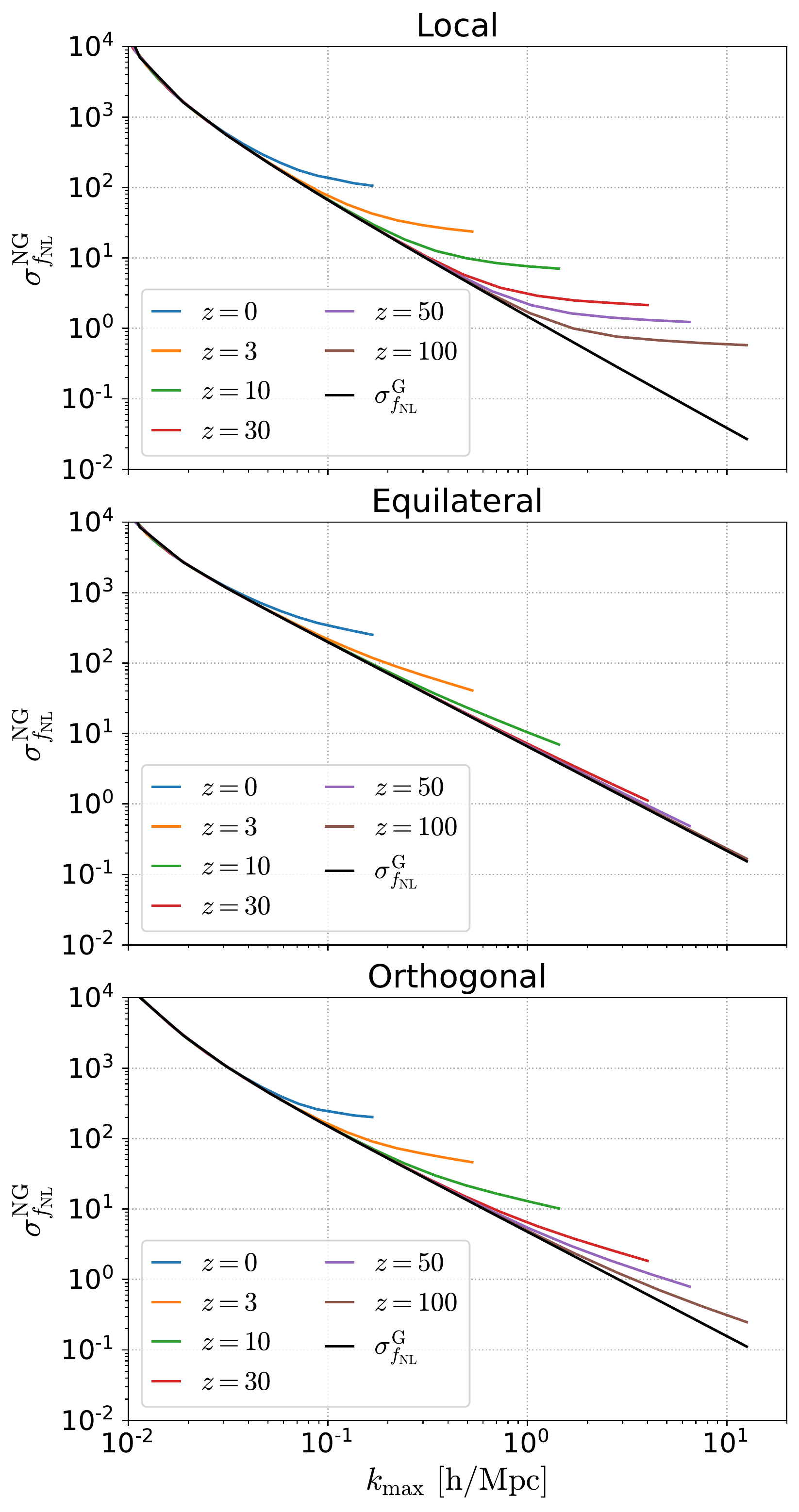}
    \caption{The estimated uncertainty on $\fnl{}$ as a function of $\kmax{}$ in the matter field at different redshifts, when including non-Gaussian covariance. The volume of the survey is taken to be 1 $(\rm Gpc/h)^3$. Each redshift is shown up to the corresponding non-linear scale $k_{\rm NL}$. Solid black is the (redshift independent) uncertainty assuming only Gaussian covariance.}
    \label{fig:snbs}
\end{figure}

To further investigate the saturation of signal-to-noise when including a non-Gaussian covariance, we plot the uncertainties on the amplitude of primordial non-Gaussianities from the matter bispectrum in real space using our model for the non-Gaussian covariance as a function of $k_{\rm max}$ in Figure \ref{fig:snbs}. All lines are plotted up to the non-linear scale $k_{\rm NL}$ for the respective redshift as given by Eq. \eqref{eq:knl}. The black line is the uncertainty computed in the approximation of only Gaussian covariance, which is redshift independent. From this visualisation, it is clear that the information content degrades when including non-Gaussian terms, even saturating before reaching the non-linear scale, especially for the local template.

\section{Details on the PUMA analysis}
\begin{figure}
    \includegraphics[scale=0.5]{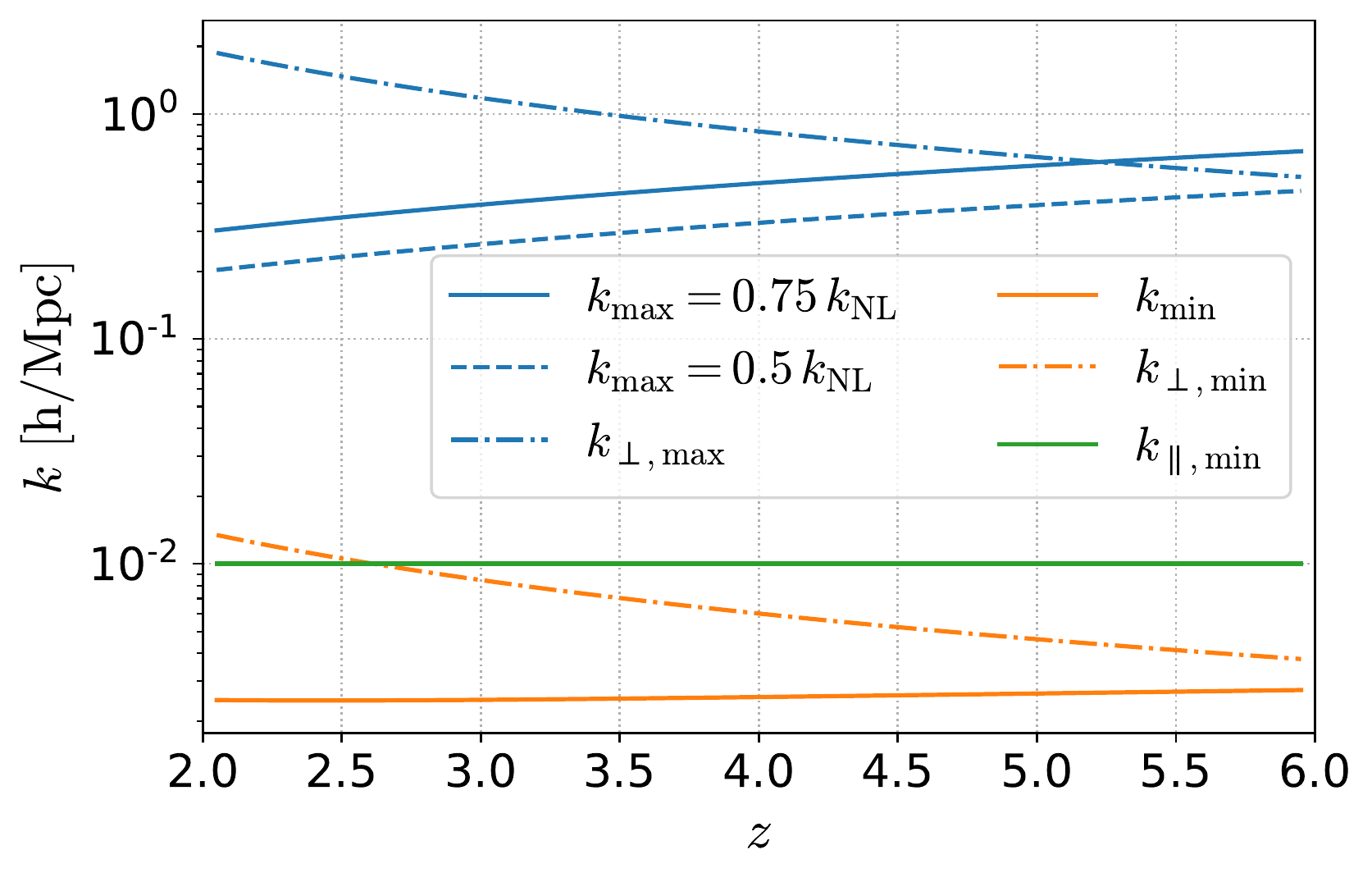}
    \caption{Smallest and largest accessible scales as a function of redshift for the PUMA survey. The smallest overall scale $k_{\rm max}$ is determined by the non-linear scale. The largest overall scale $k_{\rm min}$ is set by the volume of the redshift bin and hence depends on the choice of binning. The smallest and largest scales in the perpendicular direction, $k_{\perp ,\rm max}$ and $k_{\perp, \rm min}$ are set by properties of the experiment. Finally, we apply a foreground cut $k_{\parallel,\rm min} = 0.01$ h/Mpc in the line-of-sight direction, that effectively replaces $k_{\rm min}$, removing most dependence on the choice of binning.}
    \label{fig:PUMA_k}
\end{figure}

We perform the analysis of the impact of non-Gaussian covariance on the PUMA survey along the lines of Ref. \cite{Karagiannis:2019jjx}. We use a binning of $\Delta z = 0.1$ between $2 < z < 6$. The largest available scale is set by the volume $V_s(z)$ of the redshift bin through $k_{\text{min}}(z) = k_F(z) = 2\pi/L(z)$, where $L(z) = V_s(z)^{1/3}$, $V_s(z) = \frac{4\pi}{3}(r(z+\Delta z)^3 - r(z-\Delta z)^3)$ and $r(z)$ is the comoving distance to redshift $z$ in units of Mpc/h. 
\paragraph{Foregrounds.} 21-cm intensity mapping is complicated by foregrounds, especially on large scales in the line-of-sight direction. Therefore, the largest scale is effectively set by a foreground cut in the line-of-sight direction ($k_{\parallel,\text{min}}=  0.01 $ h/Mpc), removing much of the dependence on the choice of redshift binning. The analysis is limited to linear scales by choosing $k_{\text{max}}(z)$ to fractions of $k_{\textrm{NL}}(z)$, namely $0.5\,k_{\textrm{NL}}(z)$ and $0.75\,k_{\textrm{NL}}(z)$,  where the non-linear scale is given in Eq.~\eqref{eq:knl}. We model the foreground wedge by excluding all modes for which
\begin{eqnarray}
\label{eq:kpar}
k_\parallel < \frac{r(z)H(z)}{c(1+z)}\sin{(0.66 N_w \theta_{\rm FOV}(z))}\times k_\perp.
\end{eqnarray}
Here $H(z)$ is the Hubble parameter, $\theta_{\rm FOV}(z) = \lambda_{21}(z)/D_{\rm eff}$, $\lambda_{21}(z)$ is the redshifted 21-cm wavelength in meters and $D_{\rm eff} = (\sqrt{0.7}\times 6 \textrm{ meters})$ is the effective dish size of the PUMA setup. $N_w$ determines the severity of the foreground wedge. We apply a pessimistic wedge cut of $N_w = 3$ and a foreground cut $k_{\parallel,\rm min} = 0.01 $ h/Mpc, in order to show that the loss of constraining power persists in such a setup. Finally, the largest and smallest accessible scales in the perpendicular direction are set by:
\begin{eqnarray}
\label{eq:kper}
k_{\perp,\text{max}}(z) = \frac{2\pi D_{\rm max}}{\lambda_{21}(z)r(z)}, \hspace{0.5cm} k_{\perp,\text{min}}(z) = \frac{2\pi}{r(z)\theta_{\rm FOV}(z)},\nonumber \\ 
\end{eqnarray}
where $D_{\rm max} = 700$ meters is the largest baseline of the PUMA setup. We show all the relevant scales together in Figure \ref{fig:PUMA_k}.
\paragraph{The hydrogen power spectrum and bispectrum.} The calculation of hydrogen correlation functions in redshift-space is significantly more involved than the matter field case from the modelling point of view. Besides non-linearities in the matter field, we need to account for the biased relation between the hydrogen and matter distributions, redshift space distortions and stochasticity introduced by the discreteness effects and Poisson noise. A complete explanation of these modelling efforts can be found in \cite{Karagiannis:2019jjx}. Here we quote the hydrogen power spectrum and bispectrum for reference, defined as 
\begin{eqnarray}
\label{eq:hipow}
P_{\rm HI}(z,\bk{}) &=& P_N(z,\bk{}) +  T_b(z)^2 D_{\rm FOG}^P(z,\bk{}) \nonumber \\ 
&&\times \left[Z_1(z,\bk{})^2P_\delta^L(z,k) + P_\varepsilon(z) \right],
\end{eqnarray}
and
\begin{eqnarray}
\label{eq:hibis}
B_{\rm HI}(z,\bk{1},\bk{2},\bk{3}) &=& T_b(z)^3 \left( D_{\rm FOG}^B(z,\bk{1},\bk{2},\bk{3})\right.\nonumber \\
&\times& \Bigg[ \prod_{i=1}^3 Z_1(z,\bk{i}) B_\delta^{\rm pnG}(\bk{1},\bk{2},\bk{3})  \nonumber \\
&&+ 2 Z_1(z,\bk{1}) Z_1(z,\bk{2}) Z_2(z,\bk{1},\bk{2}) \nonumber \\ 
&& \times P_\delta^L(z,\bk{1}) P_\delta^L(z,\bk{2}) + \textrm{ 2 perm} \Bigg] \nonumber \\
&+&P_{\varepsilon \varepsilon \delta}(z)\left[\sum_{i=1}^3 Z_1(z,\bk{i})P_\delta^L(z,k_i) \right] \nonumber \\
&+& B_\varepsilon(z),
\end{eqnarray}
where $P_N$ is the instrumental noise, $T_b(z)$ is the brightness temperature of the 21-cm signal at a given redshift, $P_\varepsilon,P_{\varepsilon\varepsilon\delta},B_\varepsilon$ are stochastic noise contributions, $Z_1,Z_2$ are the first and second order redshift space kernels, $B_\delta^{\rm pnG}$ is the primordial contribution to the matter bispectrum as used in Eq. \eqref{eq:binavgB} and $D_{\rm FOG}^B$ models the Finger-Of-God dumping effect. For the explicit expressions of these quantities we refer to \cite{Karagiannis:2019jjx}. The redshift space kernels contain bias parameters $\{b_1,b_2,b_{s^2},b_\Psi, b_{\Psi\delta}\}$ as well as the linear growth rate $f$ due to redshift space distortions RSDs. 
The scale-dependent biases $\{b_\Psi,b_{\Psi\delta}\}$ can be modelled in terms of $\{b_1,b_2,\fnl{}\}$ (though see \cite{Biagetti:2016ywx} for a study of this approximation). This means primordial non-Gaussianity enters not only through $B_\delta^{\rm pnG}$ but also through the terms involving $Z_1,Z_2$, that contain the scale-dependent biases. Finally, the FOG factor is modelled using the velocity dispersion $\sigma_v$. In this work we are only interested in signal-to-noise for $\fnl{}$ coming from the hydrogen bispectrum, hence the total number of parameters including the stochastic noise contributions equals 8:
\begin{eqnarray}
\textbf{p} = \{\fnl{},b_1,b_2,b_{s^2},f,\sigma_v,P_{\varepsilon\varepsilon\delta},B_{\varepsilon}\}
\end{eqnarray}
We calculate the Fisher matrix of the 8 parameters that enter the hydrogen bispectrum, with and without non-Gaussian covariance at each redshift bin using the weighted estimator approach described above \cite{Note3}. Contrary to \cite{Karagiannis:2019jjx} we do not account for theoretical errors on the bias parameters in our analysis, which adds additional covariance (including off-diagonal) to account for uncertainties in the bias model along the lines of Ref.~\cite{Baldauf:2016sjb}. Once we have the Gaussian and non-Gaussian Fisher matrices, we marginalise over the 7 nuisance parameters by inverting the Fisher matrix at every redshift. The marginalised uncertainty for $\fnl{}$ is then given by:
\begin{eqnarray}\label{eq:marginal}
\sigma_{\fnl{}}(z) = \left(F^{-1}(z)\right)_{\fnl{}\fnl{}}^{1/2}
\end{eqnarray}
The ratio of the estimated uncertainty including non-Gaussian covariance over Gaussian covariance is shown in Figure \ref{fig:PUMA}.

\bibliographystyle{apsrev4-1}
\bibliography{bib}

%merlin.mbs apsrev4-1.bst 2010-07-25 4.21a (PWD, AO, DPC) hacked
%Control: key (0)
%Control: author (72) initials jnrlst
%Control: editor formatted (1) identically to author
%Control: production of article title (-1) disabled
%Control: page (0) single
%Control: year (1) truncated
%Control: production of eprint (0) enabled
\begin{thebibliography}{66}%
\makeatletter
\providecommand \@ifxundefined [1]{%
 \@ifx{#1\undefined}
}%
\providecommand \@ifnum [1]{%
 \ifnum #1\expandafter \@firstoftwo
 \else \expandafter \@secondoftwo
 \fi
}%
\providecommand \@ifx [1]{%
 \ifx #1\expandafter \@firstoftwo
 \else \expandafter \@secondoftwo
 \fi
}%
\providecommand \natexlab [1]{#1}%
\providecommand \enquote  [1]{``#1''}%
\providecommand \bibnamefont  [1]{#1}%
\providecommand \bibfnamefont [1]{#1}%
\providecommand \citenamefont [1]{#1}%
\providecommand \href@noop [0]{\@secondoftwo}%
\providecommand \href [0]{\begingroup \@sanitize@url \@href}%
\providecommand \@href[1]{\@@startlink{#1}\@@href}%
\providecommand \@@href[1]{\endgroup#1\@@endlink}%
\providecommand \@sanitize@url [0]{\catcode `\\12\catcode `\$12\catcode
  `\&12\catcode `\#12\catcode `\^12\catcode `\_12\catcode `\%12\relax}%
\providecommand \@@startlink[1]{}%
\providecommand \@@endlink[0]{}%
\providecommand \url  [0]{\begingroup\@sanitize@url \@url }%
\providecommand \@url [1]{\endgroup\@href {#1}{\urlprefix }}%
\providecommand \urlprefix  [0]{URL }%
\providecommand \Eprint [0]{\href }%
\providecommand \doibase [0]{http://dx.doi.org/}%
\providecommand \selectlanguage [0]{\@gobble}%
\providecommand \bibinfo  [0]{\@secondoftwo}%
\providecommand \bibfield  [0]{\@secondoftwo}%
\providecommand \translation [1]{[#1]}%
\providecommand \BibitemOpen [0]{}%
\providecommand \bibitemStop [0]{}%
\providecommand \bibitemNoStop [0]{.\EOS\space}%
\providecommand \EOS [0]{\spacefactor3000\relax}%
\providecommand \BibitemShut  [1]{\csname bibitem#1\endcsname}%
\let\auto@bib@innerbib\@empty
%</preamble>
\bibitem [{\citenamefont {Guth}(1981)}]{Guth:1980zm}%
  \BibitemOpen
  \bibfield  {author} {\bibinfo {author} {\bibfnamefont {A.~H.}\ \bibnamefont
  {Guth}},\ }\href {\doibase 10.1103/PhysRevD.23.347} {\bibfield  {journal}
  {\bibinfo  {journal} {Phys. Rev. D}\ }\textbf {\bibinfo {volume} {23}},\
  \bibinfo {pages} {347} (\bibinfo {year} {1981})}\BibitemShut {NoStop}%
\bibitem [{\citenamefont {Linde}(1982)}]{Linde:1981mu}%
  \BibitemOpen
  \bibfield  {author} {\bibinfo {author} {\bibfnamefont {A.~D.}\ \bibnamefont
  {Linde}},\ }\href {\doibase 10.1016/0370-2693(82)91219-9} {\bibfield
  {journal} {\bibinfo  {journal} {Phys. Lett. B}\ }\textbf {\bibinfo {volume}
  {108}},\ \bibinfo {pages} {389} (\bibinfo {year} {1982})}\BibitemShut
  {NoStop}%
\bibitem [{\citenamefont {Albrecht}\ and\ \citenamefont
  {Steinhardt}(1982)}]{Albrecht:1982wi}%
  \BibitemOpen
  \bibfield  {author} {\bibinfo {author} {\bibfnamefont {A.}~\bibnamefont
  {Albrecht}}\ and\ \bibinfo {author} {\bibfnamefont {P.~J.}\ \bibnamefont
  {Steinhardt}},\ }\href {\doibase 10.1103/PhysRevLett.48.1220} {\bibfield
  {journal} {\bibinfo  {journal} {Phys. Rev. Lett.}\ }\textbf {\bibinfo
  {volume} {48}},\ \bibinfo {pages} {1220} (\bibinfo {year}
  {1982})}\BibitemShut {NoStop}%
\bibitem [{\citenamefont {Akrami}\ \emph {et~al.}(2018)\citenamefont {Akrami}
  \emph {et~al.}}]{Akrami:2018odb}%
  \BibitemOpen
  \bibfield  {author} {\bibinfo {author} {\bibfnamefont {Y.}~\bibnamefont
  {Akrami}} \emph {et~al.} (\bibinfo {collaboration} {Planck}),\ }\href@noop {}
  {\  (\bibinfo {year} {2018})},\ \Eprint {http://arxiv.org/abs/1807.06211}
  {arXiv:1807.06211 [astro-ph.CO]} \BibitemShut {NoStop}%
%%CITATION = ARXIV:1807.06211;%%
\bibitem [{\citenamefont {Meerburg}\ \emph {et~al.}(2019)\citenamefont
  {Meerburg} \emph {et~al.}}]{Meerburg:2019qqi}%
  \BibitemOpen
  \bibfield  {author} {\bibinfo {author} {\bibfnamefont {P.~D.}\ \bibnamefont
  {Meerburg}} \emph {et~al.},\ }\href@noop {} {\  (\bibinfo {year} {2019})},\
  \Eprint {http://arxiv.org/abs/1903.04409} {arXiv:1903.04409 [astro-ph.CO]}
  \BibitemShut {NoStop}%
\bibitem [{\citenamefont {Ach\'ucarro}\ \emph {et~al.}(2022)\citenamefont
  {Ach\'ucarro} \emph {et~al.}}]{Achucarro:2022qrl}%
  \BibitemOpen
  \bibfield  {author} {\bibinfo {author} {\bibfnamefont {A.}~\bibnamefont
  {Ach\'ucarro}} \emph {et~al.},\ }\href@noop {} {\  (\bibinfo {year}
  {2022})},\ \Eprint {http://arxiv.org/abs/2203.08128} {arXiv:2203.08128
  [astro-ph.CO]} \BibitemShut {NoStop}%
\bibitem [{\citenamefont {Arkani-Hamed}\ and\ \citenamefont
  {Maldacena}(2015)}]{Arkani-Hamed:2015bza}%
  \BibitemOpen
  \bibfield  {author} {\bibinfo {author} {\bibfnamefont {N.}~\bibnamefont
  {Arkani-Hamed}}\ and\ \bibinfo {author} {\bibfnamefont {J.}~\bibnamefont
  {Maldacena}},\ }\href@noop {} {\  (\bibinfo {year} {2015})},\ \Eprint
  {http://arxiv.org/abs/1503.08043} {arXiv:1503.08043 [hep-th]} \BibitemShut
  {NoStop}%
%%CITATION = ARXIV:1503.08043;%%
\bibitem [{\citenamefont {Takahashi}\ \emph {et~al.}(2011)\citenamefont
  {Takahashi}, \citenamefont {Yoshida}, \citenamefont {Takada}, \citenamefont
  {Matsubara}, \citenamefont {Sugiyama}, \citenamefont {Kayo}, \citenamefont
  {Nishimichi}, \citenamefont {Saito},\ and\ \citenamefont
  {Taruya}}]{Takahashi:2009ty}%
  \BibitemOpen
  \bibfield  {author} {\bibinfo {author} {\bibfnamefont {R.}~\bibnamefont
  {Takahashi}}, \bibinfo {author} {\bibfnamefont {N.}~\bibnamefont {Yoshida}},
  \bibinfo {author} {\bibfnamefont {M.}~\bibnamefont {Takada}}, \bibinfo
  {author} {\bibfnamefont {T.}~\bibnamefont {Matsubara}}, \bibinfo {author}
  {\bibfnamefont {N.}~\bibnamefont {Sugiyama}}, \bibinfo {author}
  {\bibfnamefont {I.}~\bibnamefont {Kayo}}, \bibinfo {author} {\bibfnamefont
  {T.}~\bibnamefont {Nishimichi}}, \bibinfo {author} {\bibfnamefont
  {S.}~\bibnamefont {Saito}}, \ and\ \bibinfo {author} {\bibfnamefont
  {A.}~\bibnamefont {Taruya}},\ }\href {\doibase 10.1088/0004-637X/726/1/7}
  {\bibfield  {journal} {\bibinfo  {journal} {Astrophys. J.}\ }\textbf
  {\bibinfo {volume} {726}},\ \bibinfo {pages} {7} (\bibinfo {year} {2011})},\
  \Eprint {http://arxiv.org/abs/0912.1381} {arXiv:0912.1381 [astro-ph.CO]}
  \BibitemShut {NoStop}%
\bibitem [{\citenamefont {Chan}\ and\ \citenamefont
  {Blot}(2017)}]{Chan:2016ehg}%
  \BibitemOpen
  \bibfield  {author} {\bibinfo {author} {\bibfnamefont {K.~C.}\ \bibnamefont
  {Chan}}\ and\ \bibinfo {author} {\bibfnamefont {L.}~\bibnamefont {Blot}},\
  }\href {\doibase 10.1103/PhysRevD.96.023528} {\bibfield  {journal} {\bibinfo
  {journal} {Phys. Rev. D}\ }\textbf {\bibinfo {volume} {96}},\ \bibinfo
  {pages} {023528} (\bibinfo {year} {2017})},\ \Eprint
  {http://arxiv.org/abs/1610.06585} {arXiv:1610.06585 [astro-ph.CO]}
  \BibitemShut {NoStop}%
\bibitem [{\citenamefont {Chan}\ \emph {et~al.}(2018)\citenamefont {Chan},
  \citenamefont {Moradinezhad~Dizgah},\ and\ \citenamefont
  {Nore\~na}}]{Chan:2017fiv}%
  \BibitemOpen
  \bibfield  {author} {\bibinfo {author} {\bibfnamefont {K.~C.}\ \bibnamefont
  {Chan}}, \bibinfo {author} {\bibfnamefont {A.}~\bibnamefont
  {Moradinezhad~Dizgah}}, \ and\ \bibinfo {author} {\bibfnamefont
  {J.}~\bibnamefont {Nore\~na}},\ }\href {\doibase 10.1103/PhysRevD.97.043532}
  {\bibfield  {journal} {\bibinfo  {journal} {Phys. Rev. D}\ }\textbf {\bibinfo
  {volume} {97}},\ \bibinfo {pages} {043532} (\bibinfo {year} {2018})},\
  \Eprint {http://arxiv.org/abs/1709.02473} {arXiv:1709.02473 [astro-ph.CO]}
  \BibitemShut {NoStop}%
\bibitem [{\citenamefont {Wadekar}\ and\ \citenamefont
  {Scoccimarro}(2020)}]{Wadekar:2019rdu}%
  \BibitemOpen
  \bibfield  {author} {\bibinfo {author} {\bibfnamefont {D.}~\bibnamefont
  {Wadekar}}\ and\ \bibinfo {author} {\bibfnamefont {R.}~\bibnamefont
  {Scoccimarro}},\ }\href {\doibase 10.1103/PhysRevD.102.123517} {\bibfield
  {journal} {\bibinfo  {journal} {Phys. Rev. D}\ }\textbf {\bibinfo {volume}
  {102}},\ \bibinfo {pages} {123517} (\bibinfo {year} {2020})},\ \Eprint
  {http://arxiv.org/abs/1910.02914} {arXiv:1910.02914 [astro-ph.CO]}
  \BibitemShut {NoStop}%
\bibitem [{\citenamefont {Barreira}(2019)}]{Barreira:2019icq}%
  \BibitemOpen
  \bibfield  {author} {\bibinfo {author} {\bibfnamefont {A.}~\bibnamefont
  {Barreira}},\ }\href {\doibase 10.1088/1475-7516/2019/03/008} {\bibfield
  {journal} {\bibinfo  {journal} {JCAP}\ }\textbf {\bibinfo {volume} {03}},\
  \bibinfo {pages} {008} (\bibinfo {year} {2019})},\ \Eprint
  {http://arxiv.org/abs/1901.01243} {arXiv:1901.01243 [astro-ph.CO]}
  \BibitemShut {NoStop}%
\bibitem [{\citenamefont {Gualdi}\ and\ \citenamefont
  {Verde}(2020)}]{Gualdi:2020ymf}%
  \BibitemOpen
  \bibfield  {author} {\bibinfo {author} {\bibfnamefont {D.}~\bibnamefont
  {Gualdi}}\ and\ \bibinfo {author} {\bibfnamefont {L.}~\bibnamefont {Verde}},\
  }\href {\doibase 10.1088/1475-7516/2020/06/041} {\bibfield  {journal}
  {\bibinfo  {journal} {JCAP}\ }\textbf {\bibinfo {volume} {06}},\ \bibinfo
  {pages} {041} (\bibinfo {year} {2020})},\ \Eprint
  {http://arxiv.org/abs/2003.12075} {arXiv:2003.12075 [astro-ph.CO]}
  \BibitemShut {NoStop}%
\bibitem [{\citenamefont {Oddo}\ \emph {et~al.}(2021)\citenamefont {Oddo},
  \citenamefont {Rizzo}, \citenamefont {Sefusatti}, \citenamefont {Porciani},\
  and\ \citenamefont {Monaco}}]{Oddo:2021iwq}%
  \BibitemOpen
  \bibfield  {author} {\bibinfo {author} {\bibfnamefont {A.}~\bibnamefont
  {Oddo}}, \bibinfo {author} {\bibfnamefont {F.}~\bibnamefont {Rizzo}},
  \bibinfo {author} {\bibfnamefont {E.}~\bibnamefont {Sefusatti}}, \bibinfo
  {author} {\bibfnamefont {C.}~\bibnamefont {Porciani}}, \ and\ \bibinfo
  {author} {\bibfnamefont {P.}~\bibnamefont {Monaco}},\ }\href {\doibase
  10.1088/1475-7516/2021/11/038} {\bibfield  {journal} {\bibinfo  {journal}
  {JCAP}\ }\textbf {\bibinfo {volume} {11}},\ \bibinfo {pages} {038} (\bibinfo
  {year} {2021})},\ \Eprint {http://arxiv.org/abs/2108.03204} {arXiv:2108.03204
  [astro-ph.CO]} \BibitemShut {NoStop}%
\bibitem [{\citenamefont {Barreira}(2022)}]{Barreira:2021ueb}%
  \BibitemOpen
  \bibfield  {author} {\bibinfo {author} {\bibfnamefont {A.}~\bibnamefont
  {Barreira}},\ }\href {\doibase 10.1088/1475-7516/2022/01/033} {\bibfield
  {journal} {\bibinfo  {journal} {JCAP}\ }\textbf {\bibinfo {volume} {01}},\
  \bibinfo {pages} {033} (\bibinfo {year} {2022})},\ \Eprint
  {http://arxiv.org/abs/2107.06887} {arXiv:2107.06887 [astro-ph.CO]}
  \BibitemShut {NoStop}%
\bibitem [{\citenamefont {Biagetti}\ \emph {et~al.}(2021)\citenamefont
  {Biagetti}, \citenamefont {Castiblanco}, \citenamefont {Nore\~na},\ and\
  \citenamefont {Sefusatti}}]{Biagetti:2021tua}%
  \BibitemOpen
  \bibfield  {author} {\bibinfo {author} {\bibfnamefont {M.}~\bibnamefont
  {Biagetti}}, \bibinfo {author} {\bibfnamefont {L.}~\bibnamefont
  {Castiblanco}}, \bibinfo {author} {\bibfnamefont {J.}~\bibnamefont
  {Nore\~na}}, \ and\ \bibinfo {author} {\bibfnamefont {E.}~\bibnamefont
  {Sefusatti}},\ }\href@noop {} {\  (\bibinfo {year} {2021})},\ \Eprint
  {http://arxiv.org/abs/2111.05887} {arXiv:2111.05887 [astro-ph.CO]}
  \BibitemShut {NoStop}%
\bibitem [{\citenamefont {Rizzo}\ \emph {et~al.}(2022)\citenamefont {Rizzo},
  \citenamefont {Moretti}, \citenamefont {Pardede}, \citenamefont {Eggemeier},
  \citenamefont {Oddo}, \citenamefont {Sefusatti}, \citenamefont {Porciani},\
  and\ \citenamefont {Monaco}}]{Rizzo:2022lmh}%
  \BibitemOpen
  \bibfield  {author} {\bibinfo {author} {\bibfnamefont {F.}~\bibnamefont
  {Rizzo}}, \bibinfo {author} {\bibfnamefont {C.}~\bibnamefont {Moretti}},
  \bibinfo {author} {\bibfnamefont {K.}~\bibnamefont {Pardede}}, \bibinfo
  {author} {\bibfnamefont {A.}~\bibnamefont {Eggemeier}}, \bibinfo {author}
  {\bibfnamefont {A.}~\bibnamefont {Oddo}}, \bibinfo {author} {\bibfnamefont
  {E.}~\bibnamefont {Sefusatti}}, \bibinfo {author} {\bibfnamefont
  {C.}~\bibnamefont {Porciani}}, \ and\ \bibinfo {author} {\bibfnamefont
  {P.}~\bibnamefont {Monaco}},\ }\href@noop {} {\  (\bibinfo {year} {2022})},\
  \Eprint {http://arxiv.org/abs/2204.13628} {arXiv:2204.13628 [astro-ph.CO]}
  \BibitemShut {NoStop}%
\bibitem [{\citenamefont {Mu\~noz}\ \emph {et~al.}(2015)\citenamefont
  {Mu\~noz}, \citenamefont {Ali-Ha\"\i{}moud},\ and\ \citenamefont
  {Kamionkowski}}]{Munoz:2015eqa}%
  \BibitemOpen
  \bibfield  {author} {\bibinfo {author} {\bibfnamefont {J.~B.}\ \bibnamefont
  {Mu\~noz}}, \bibinfo {author} {\bibfnamefont {Y.}~\bibnamefont
  {Ali-Ha\"\i{}moud}}, \ and\ \bibinfo {author} {\bibfnamefont
  {M.}~\bibnamefont {Kamionkowski}},\ }\href {\doibase
  10.1103/PhysRevD.92.083508} {\bibfield  {journal} {\bibinfo  {journal} {Phys.
  Rev. D}\ }\textbf {\bibinfo {volume} {92}},\ \bibinfo {pages} {083508}
  (\bibinfo {year} {2015})},\ \Eprint {http://arxiv.org/abs/1506.04152}
  {arXiv:1506.04152 [astro-ph.CO]} \BibitemShut {NoStop}%
\bibitem [{\citenamefont {Chen}\ \emph {et~al.}(2016)\citenamefont {Chen},
  \citenamefont {Meerburg},\ and\ \citenamefont {M\"unchmeyer}}]{Chen:2016zuu}%
  \BibitemOpen
  \bibfield  {author} {\bibinfo {author} {\bibfnamefont {X.}~\bibnamefont
  {Chen}}, \bibinfo {author} {\bibfnamefont {P.~D.}\ \bibnamefont {Meerburg}},
  \ and\ \bibinfo {author} {\bibfnamefont {M.}~\bibnamefont {M\"unchmeyer}},\
  }\href {\doibase 10.1088/1475-7516/2016/09/023} {\bibfield  {journal}
  {\bibinfo  {journal} {JCAP}\ }\textbf {\bibinfo {volume} {09}},\ \bibinfo
  {pages} {023} (\bibinfo {year} {2016})},\ \Eprint
  {http://arxiv.org/abs/1605.09364} {arXiv:1605.09364 [astro-ph.CO]}
  \BibitemShut {NoStop}%
\bibitem [{\citenamefont {Meerburg}\ \emph {et~al.}(2017)\citenamefont
  {Meerburg}, \citenamefont {M\"unchmeyer}, \citenamefont {Mu\~noz},\ and\
  \citenamefont {Chen}}]{Meerburg:2016zdz}%
  \BibitemOpen
  \bibfield  {author} {\bibinfo {author} {\bibfnamefont {P.~D.}\ \bibnamefont
  {Meerburg}}, \bibinfo {author} {\bibfnamefont {M.}~\bibnamefont
  {M\"unchmeyer}}, \bibinfo {author} {\bibfnamefont {J.~B.}\ \bibnamefont
  {Mu\~noz}}, \ and\ \bibinfo {author} {\bibfnamefont {X.}~\bibnamefont
  {Chen}},\ }\href {\doibase 10.1088/1475-7516/2017/03/050} {\bibfield
  {journal} {\bibinfo  {journal} {JCAP}\ }\textbf {\bibinfo {volume} {03}},\
  \bibinfo {pages} {050} (\bibinfo {year} {2017})},\ \Eprint
  {http://arxiv.org/abs/1610.06559} {arXiv:1610.06559 [astro-ph.CO]}
  \BibitemShut {NoStop}%
\bibitem [{\citenamefont {Karagiannis}\ \emph {et~al.}(2020)\citenamefont
  {Karagiannis}, \citenamefont {Slosar},\ and\ \citenamefont
  {Liguori}}]{Karagiannis:2019jjx}%
  \BibitemOpen
  \bibfield  {author} {\bibinfo {author} {\bibfnamefont {D.}~\bibnamefont
  {Karagiannis}}, \bibinfo {author} {\bibfnamefont {A.}~\bibnamefont {Slosar}},
  \ and\ \bibinfo {author} {\bibfnamefont {M.}~\bibnamefont {Liguori}},\ }\href
  {\doibase 10.1088/1475-7516/2020/11/052} {\bibfield  {journal} {\bibinfo
  {journal} {JCAP}\ }\textbf {\bibinfo {volume} {11}},\ \bibinfo {pages} {052}
  (\bibinfo {year} {2020})},\ \Eprint {http://arxiv.org/abs/1911.03964}
  {arXiv:1911.03964 [astro-ph.CO]} \BibitemShut {NoStop}%
\bibitem [{\citenamefont {Fl\"oss}\ \emph {et~al.}(2022)\citenamefont
  {Fl\"oss}, \citenamefont {de~Wild}, \citenamefont {Meerburg},\ and\
  \citenamefont {Koopmans}}]{Floss:2022grj}%
  \BibitemOpen
  \bibfield  {author} {\bibinfo {author} {\bibfnamefont {T.}~\bibnamefont
  {Fl\"oss}}, \bibinfo {author} {\bibfnamefont {T.}~\bibnamefont {de~Wild}},
  \bibinfo {author} {\bibfnamefont {P.~D.}\ \bibnamefont {Meerburg}}, \ and\
  \bibinfo {author} {\bibfnamefont {L.~V.~E.}\ \bibnamefont {Koopmans}},\
  }\href@noop {} {\  (\bibinfo {year} {2022})},\ \Eprint
  {http://arxiv.org/abs/2201.08843} {arXiv:2201.08843 [astro-ph.CO]}
  \BibitemShut {NoStop}%
\bibitem [{\citenamefont {Yamauchi}(2022)}]{Yamauchi:2022fri}%
  \BibitemOpen
  \bibfield  {author} {\bibinfo {author} {\bibfnamefont {D.}~\bibnamefont
  {Yamauchi}},\ }\href@noop {} {\  (\bibinfo {year} {2022})},\ \Eprint
  {http://arxiv.org/abs/2203.15599} {arXiv:2203.15599 [astro-ph.CO]}
  \BibitemShut {NoStop}%
\bibitem [{\citenamefont {Karagiannis}\ \emph {et~al.}(2022)\citenamefont
  {Karagiannis}, \citenamefont {Maartens},\ and\ \citenamefont
  {Randrianjanahary}}]{Karagiannis:2022ylq}%
  \BibitemOpen
  \bibfield  {author} {\bibinfo {author} {\bibfnamefont {D.}~\bibnamefont
  {Karagiannis}}, \bibinfo {author} {\bibfnamefont {R.}~\bibnamefont
  {Maartens}}, \ and\ \bibinfo {author} {\bibfnamefont {L.}~\bibnamefont
  {Randrianjanahary}},\ }\href@noop {} {\  (\bibinfo {year} {2022})},\ \Eprint
  {http://arxiv.org/abs/2206.07747} {arXiv:2206.07747 [astro-ph.CO]}
  \BibitemShut {NoStop}%
\bibitem [{\citenamefont {Ansari}\ \emph {et~al.}(2018)\citenamefont {Ansari}
  \emph {et~al.}}]{CosmicVisions21cm:2018rfq}%
  \BibitemOpen
  \bibfield  {author} {\bibinfo {author} {\bibfnamefont {R.}~\bibnamefont
  {Ansari}} \emph {et~al.} (\bibinfo {collaboration} {Cosmic Visions 21 cm}),\
  }\href@noop {} {\  (\bibinfo {year} {2018})},\ \Eprint
  {http://arxiv.org/abs/1810.09572} {arXiv:1810.09572 [astro-ph.CO]}
  \BibitemShut {NoStop}%
\bibitem [{\citenamefont {Villaescusa-Navarro}\ \emph
  {et~al.}(2020)\citenamefont {Villaescusa-Navarro} \emph
  {et~al.}}]{Villaescusa-Navarro:2019bje}%
  \BibitemOpen
  \bibfield  {author} {\bibinfo {author} {\bibfnamefont {F.}~\bibnamefont
  {Villaescusa-Navarro}} \emph {et~al.},\ }\href {\doibase
  10.3847/1538-4365/ab9d82} {\bibfield  {journal} {\bibinfo  {journal}
  {Astrophys. J. Suppl.}\ }\textbf {\bibinfo {volume} {250}},\ \bibinfo {pages}
  {2} (\bibinfo {year} {2020})},\ \Eprint {http://arxiv.org/abs/1909.05273}
  {arXiv:1909.05273 [astro-ph.CO]} \BibitemShut {NoStop}%
\bibitem [{\citenamefont {Aghanim}\ \emph {et~al.}(2020)\citenamefont {Aghanim}
  \emph {et~al.}}]{Planck:2018vyg}%
  \BibitemOpen
  \bibfield  {author} {\bibinfo {author} {\bibfnamefont {N.}~\bibnamefont
  {Aghanim}} \emph {et~al.} (\bibinfo {collaboration} {Planck}),\ }\href
  {\doibase 10.1051/0004-6361/201833910} {\bibfield  {journal} {\bibinfo
  {journal} {Astron. Astrophys.}\ }\textbf {\bibinfo {volume} {641}},\ \bibinfo
  {pages} {A6} (\bibinfo {year} {2020})},\ \bibinfo {note} {[Erratum:
  Astron.Astrophys. 652, C4 (2021)]},\ \Eprint
  {http://arxiv.org/abs/1807.06209} {arXiv:1807.06209 [astro-ph.CO]}
  \BibitemShut {NoStop}%
\bibitem [{\citenamefont {Ade}\ \emph {et~al.}(2016)\citenamefont {Ade} \emph
  {et~al.}}]{Planck:2015fie}%
  \BibitemOpen
  \bibfield  {author} {\bibinfo {author} {\bibfnamefont {P.~A.~R.}\
  \bibnamefont {Ade}} \emph {et~al.} (\bibinfo {collaboration} {Planck}),\
  }\href {\doibase 10.1051/0004-6361/201525830} {\bibfield  {journal} {\bibinfo
   {journal} {Astron. Astrophys.}\ }\textbf {\bibinfo {volume} {594}},\
  \bibinfo {pages} {A13} (\bibinfo {year} {2016})},\ \Eprint
  {http://arxiv.org/abs/1502.01589} {arXiv:1502.01589 [astro-ph.CO]}
  \BibitemShut {NoStop}%
\bibitem [{\citenamefont {Lee}\ \emph {et~al.}(2016)\citenamefont {Lee},
  \citenamefont {Baumann},\ and\ \citenamefont {Pimentel}}]{Lee:2016vti}%
  \BibitemOpen
  \bibfield  {author} {\bibinfo {author} {\bibfnamefont {H.}~\bibnamefont
  {Lee}}, \bibinfo {author} {\bibfnamefont {D.}~\bibnamefont {Baumann}}, \ and\
  \bibinfo {author} {\bibfnamefont {G.~L.}\ \bibnamefont {Pimentel}},\ }\href
  {\doibase 10.1007/JHEP12(2016)040} {\bibfield  {journal} {\bibinfo  {journal}
  {JHEP}\ }\textbf {\bibinfo {volume} {12}},\ \bibinfo {pages} {040} (\bibinfo
  {year} {2016})},\ \Eprint {http://arxiv.org/abs/1607.03735} {arXiv:1607.03735
  [hep-th]} \BibitemShut {NoStop}%
\bibitem [{\citenamefont {Cheung}\ \emph {et~al.}(2008)\citenamefont {Cheung},
  \citenamefont {Creminelli}, \citenamefont {Fitzpatrick}, \citenamefont
  {Kaplan},\ and\ \citenamefont {Senatore}}]{Cheung:2007st}%
  \BibitemOpen
  \bibfield  {author} {\bibinfo {author} {\bibfnamefont {C.}~\bibnamefont
  {Cheung}}, \bibinfo {author} {\bibfnamefont {P.}~\bibnamefont {Creminelli}},
  \bibinfo {author} {\bibfnamefont {A.~L.}\ \bibnamefont {Fitzpatrick}},
  \bibinfo {author} {\bibfnamefont {J.}~\bibnamefont {Kaplan}}, \ and\ \bibinfo
  {author} {\bibfnamefont {L.}~\bibnamefont {Senatore}},\ }\href {\doibase
  10.1088/1126-6708/2008/03/014} {\bibfield  {journal} {\bibinfo  {journal}
  {JHEP}\ }\textbf {\bibinfo {volume} {03}},\ \bibinfo {pages} {014} (\bibinfo
  {year} {2008})},\ \Eprint {http://arxiv.org/abs/0709.0293} {arXiv:0709.0293
  [hep-th]} \BibitemShut {NoStop}%
\bibitem [{\citenamefont {Bernardeau}\ \emph {et~al.}(2002)\citenamefont
  {Bernardeau}, \citenamefont {Colombi}, \citenamefont {Gaztanaga},\ and\
  \citenamefont {Scoccimarro}}]{Bernardeau:2001qr}%
  \BibitemOpen
  \bibfield  {author} {\bibinfo {author} {\bibfnamefont {F.}~\bibnamefont
  {Bernardeau}}, \bibinfo {author} {\bibfnamefont {S.}~\bibnamefont {Colombi}},
  \bibinfo {author} {\bibfnamefont {E.}~\bibnamefont {Gaztanaga}}, \ and\
  \bibinfo {author} {\bibfnamefont {R.}~\bibnamefont {Scoccimarro}},\ }\href
  {\doibase 10.1016/S0370-1573(02)00135-7} {\bibfield  {journal} {\bibinfo
  {journal} {Phys. Rept.}\ }\textbf {\bibinfo {volume} {367}},\ \bibinfo
  {pages} {1} (\bibinfo {year} {2002})},\ \Eprint
  {http://arxiv.org/abs/astro-ph/0112551} {arXiv:astro-ph/0112551} \BibitemShut
  {NoStop}%
\bibitem [{Note1()}]{Note1}%
  \BibitemOpen
  \bibinfo {note} {Other definitions have been considered in the literature,
  e.g. \cite {Tomlinson:2022xud} studies the non-linear scale for the
  bispectrum specifically. The precise definition of $k_{\protect \rm NL}$ does
  not qualitatively change the results of this paper.}\BibitemShut {Stop}%
\bibitem [{\citenamefont {Angulo}\ and\ \citenamefont
  {Hahn}(2021)}]{Angulo:2021kes}%
  \BibitemOpen
  \bibfield  {author} {\bibinfo {author} {\bibfnamefont {R.~E.}\ \bibnamefont
  {Angulo}}\ and\ \bibinfo {author} {\bibfnamefont {O.}~\bibnamefont {Hahn}},\
  }\href {\doibase 10.1007/s41115-021-00013-z} {\  (\bibinfo {year} {2021}),\
  10.1007/s41115-021-00013-z},\ \Eprint {http://arxiv.org/abs/2112.05165}
  {arXiv:2112.05165 [astro-ph.CO]} \BibitemShut {NoStop}%
\bibitem [{\citenamefont {Michaux}\ \emph {et~al.}(2020)\citenamefont
  {Michaux}, \citenamefont {Hahn}, \citenamefont {Rampf},\ and\ \citenamefont
  {Angulo}}]{Michaux:2020yis}%
  \BibitemOpen
  \bibfield  {author} {\bibinfo {author} {\bibfnamefont {M.}~\bibnamefont
  {Michaux}}, \bibinfo {author} {\bibfnamefont {O.}~\bibnamefont {Hahn}},
  \bibinfo {author} {\bibfnamefont {C.}~\bibnamefont {Rampf}}, \ and\ \bibinfo
  {author} {\bibfnamefont {R.~E.}\ \bibnamefont {Angulo}},\ }\href {\doibase
  10.1093/mnras/staa3149} {\bibfield  {journal} {\bibinfo  {journal} {Mon. Not.
  Roy. Astron. Soc.}\ }\textbf {\bibinfo {volume} {500}},\ \bibinfo {pages}
  {663} (\bibinfo {year} {2020})},\ \Eprint {http://arxiv.org/abs/2008.09588}
  {arXiv:2008.09588 [astro-ph.CO]} \BibitemShut {NoStop}%
\bibitem [{\citenamefont {Scoccimarro}\ \emph {et~al.}(1998)\citenamefont
  {Scoccimarro}, \citenamefont {Colombi}, \citenamefont {Fry}, \citenamefont
  {Frieman}, \citenamefont {Hivon},\ and\ \citenamefont
  {Melott}}]{Scoccimarro:1997st}%
  \BibitemOpen
  \bibfield  {author} {\bibinfo {author} {\bibfnamefont {R.}~\bibnamefont
  {Scoccimarro}}, \bibinfo {author} {\bibfnamefont {S.}~\bibnamefont
  {Colombi}}, \bibinfo {author} {\bibfnamefont {J.~N.}\ \bibnamefont {Fry}},
  \bibinfo {author} {\bibfnamefont {J.~A.}\ \bibnamefont {Frieman}}, \bibinfo
  {author} {\bibfnamefont {E.}~\bibnamefont {Hivon}}, \ and\ \bibinfo {author}
  {\bibfnamefont {A.}~\bibnamefont {Melott}},\ }\href {\doibase 10.1086/305399}
  {\bibfield  {journal} {\bibinfo  {journal} {Astrophys. J.}\ }\textbf
  {\bibinfo {volume} {496}},\ \bibinfo {pages} {586} (\bibinfo {year}
  {1998})},\ \Eprint {http://arxiv.org/abs/astro-ph/9704075}
  {arXiv:astro-ph/9704075} \BibitemShut {NoStop}%
\bibitem [{Note2()}]{Note2}%
  \BibitemOpen
  \bibinfo {note} {We also measure the power spectrum, since, as we show below,
  it enters in the calculation of the covariance. The estimator of the power
  spectrum is \begin {equation}\label {eq:estp} \protect \hat P(k) \equiv
  \protect \frac {k^3_F}{N_k}\DOTSB \sum@ \slimits@ _{\protect \mathbf {q} \in
  k} \delta _{\protect \mathbf {q}}\protect \,\delta _{-\protect \mathbf {q}},
  \end {equation} where $N_k$ gives the number of modes in each
  k-bin.}\BibitemShut {Stop}%
\bibitem [{Note3()}]{Note3}%
  \BibitemOpen
  \bibinfo {note} {The approximate equality indicates the thin shell
  approximation.}\BibitemShut {Stop}%
\bibitem [{Note4()}]{Note4}%
  \BibitemOpen
  \bibinfo {note} {This is very similar to how lensing-induced covariance
  mostly affects measurements of the local bispectrum in the CMB \cite
  {Coulton:2019odk}}\BibitemShut {NoStop}%
\bibitem [{\citenamefont {Desjacques}\ \emph {et~al.}(2018)\citenamefont
  {Desjacques}, \citenamefont {Jeong},\ and\ \citenamefont
  {Schmidt}}]{Desjacques:2016bnm}%
  \BibitemOpen
  \bibfield  {author} {\bibinfo {author} {\bibfnamefont {V.}~\bibnamefont
  {Desjacques}}, \bibinfo {author} {\bibfnamefont {D.}~\bibnamefont {Jeong}}, \
  and\ \bibinfo {author} {\bibfnamefont {F.}~\bibnamefont {Schmidt}},\ }\href
  {\doibase 10.1016/j.physrep.2017.12.002} {\bibfield  {journal} {\bibinfo
  {journal} {Phys. Rept.}\ }\textbf {\bibinfo {volume} {733}},\ \bibinfo
  {pages} {1} (\bibinfo {year} {2018})},\ \Eprint
  {http://arxiv.org/abs/1611.09787} {arXiv:1611.09787 [astro-ph.CO]}
  \BibitemShut {NoStop}%
\bibitem [{\citenamefont {Dalal}\ \emph {et~al.}(2008)\citenamefont {Dalal},
  \citenamefont {Dore}, \citenamefont {Huterer},\ and\ \citenamefont
  {Shirokov}}]{Dalal:2007cu}%
  \BibitemOpen
  \bibfield  {author} {\bibinfo {author} {\bibfnamefont {N.}~\bibnamefont
  {Dalal}}, \bibinfo {author} {\bibfnamefont {O.}~\bibnamefont {Dore}},
  \bibinfo {author} {\bibfnamefont {D.}~\bibnamefont {Huterer}}, \ and\
  \bibinfo {author} {\bibfnamefont {A.}~\bibnamefont {Shirokov}},\ }\href
  {\doibase 10.1103/PhysRevD.77.123514} {\bibfield  {journal} {\bibinfo
  {journal} {Phys. Rev. D}\ }\textbf {\bibinfo {volume} {77}},\ \bibinfo
  {pages} {123514} (\bibinfo {year} {2008})},\ \Eprint
  {http://arxiv.org/abs/0710.4560} {arXiv:0710.4560 [astro-ph]} \BibitemShut
  {NoStop}%
\bibitem [{\citenamefont {Matarrese}\ and\ \citenamefont
  {Verde}(2008)}]{Matarrese:2008nc}%
  \BibitemOpen
  \bibfield  {author} {\bibinfo {author} {\bibfnamefont {S.}~\bibnamefont
  {Matarrese}}\ and\ \bibinfo {author} {\bibfnamefont {L.}~\bibnamefont
  {Verde}},\ }\href {\doibase 10.1086/587840} {\bibfield  {journal} {\bibinfo
  {journal} {Astrophys. J. Lett.}\ }\textbf {\bibinfo {volume} {677}},\
  \bibinfo {pages} {L77} (\bibinfo {year} {2008})},\ \Eprint
  {http://arxiv.org/abs/0801.4826} {arXiv:0801.4826 [astro-ph]} \BibitemShut
  {NoStop}%
\bibitem [{\citenamefont {Slosar}\ \emph {et~al.}(2008)\citenamefont {Slosar},
  \citenamefont {Hirata}, \citenamefont {Seljak}, \citenamefont {Ho},\ and\
  \citenamefont {Padmanabhan}}]{Slosar:2008hx}%
  \BibitemOpen
  \bibfield  {author} {\bibinfo {author} {\bibfnamefont {A.}~\bibnamefont
  {Slosar}}, \bibinfo {author} {\bibfnamefont {C.}~\bibnamefont {Hirata}},
  \bibinfo {author} {\bibfnamefont {U.}~\bibnamefont {Seljak}}, \bibinfo
  {author} {\bibfnamefont {S.}~\bibnamefont {Ho}}, \ and\ \bibinfo {author}
  {\bibfnamefont {N.}~\bibnamefont {Padmanabhan}},\ }\href {\doibase
  10.1088/1475-7516/2008/08/031} {\bibfield  {journal} {\bibinfo  {journal}
  {JCAP}\ }\textbf {\bibinfo {volume} {08}},\ \bibinfo {pages} {031} (\bibinfo
  {year} {2008})},\ \Eprint {http://arxiv.org/abs/0805.3580} {arXiv:0805.3580
  [astro-ph]} \BibitemShut {NoStop}%
\bibitem [{\citenamefont {Biagetti}(2019)}]{Biagetti:2019bnp}%
  \BibitemOpen
  \bibfield  {author} {\bibinfo {author} {\bibfnamefont {M.}~\bibnamefont
  {Biagetti}},\ }\href {\doibase 10.3390/galaxies7030071} {\bibfield  {journal}
  {\bibinfo  {journal} {Galaxies}\ }\textbf {\bibinfo {volume} {7}},\ \bibinfo
  {pages} {71} (\bibinfo {year} {2019})},\ \Eprint
  {http://arxiv.org/abs/1906.12244} {arXiv:1906.12244 [astro-ph.CO]}
  \BibitemShut {NoStop}%
\bibitem [{Note5()}]{Note5}%
  \BibitemOpen
  \bibinfo {note} {We have confirmed that our forecasts, using very similar
  assumption about the PUMA survey, result in forecasts on $\sigma (f_{\protect
  \rm NL})$ that are consistent with those presented in Refs.~\cite
  {Karagiannis:2019jjx,Sailer:2021yzm} when neglecting non-Gaussian
  covariance}\BibitemShut {NoStop}%
\bibitem [{\citenamefont {Castorina}\ and\ \citenamefont
  {Villaescusa-Navarro}(2017)}]{Castorina:2016bfm}%
  \BibitemOpen
  \bibfield  {author} {\bibinfo {author} {\bibfnamefont {E.}~\bibnamefont
  {Castorina}}\ and\ \bibinfo {author} {\bibfnamefont {F.}~\bibnamefont
  {Villaescusa-Navarro}},\ }\href {\doibase 10.1093/mnras/stx1599} {\bibfield
  {journal} {\bibinfo  {journal} {Mon. Not. Roy. Astron. Soc.}\ }\textbf
  {\bibinfo {volume} {471}},\ \bibinfo {pages} {1788} (\bibinfo {year}
  {2017})},\ \Eprint {http://arxiv.org/abs/1609.05157} {arXiv:1609.05157
  [astro-ph.CO]} \BibitemShut {NoStop}%
\bibitem [{\citenamefont {de~Putter}(2018)}]{dePutter:2018jqk}%
  \BibitemOpen
  \bibfield  {author} {\bibinfo {author} {\bibfnamefont {R.}~\bibnamefont
  {de~Putter}},\ }\href@noop {} {\  (\bibinfo {year} {2018})},\ \Eprint
  {http://arxiv.org/abs/1802.06762} {arXiv:1802.06762 [astro-ph.CO]}
  \BibitemShut {NoStop}%
\bibitem [{\citenamefont {Schlegel}\ \emph {et~al.}(2019)\citenamefont
  {Schlegel} \emph {et~al.}}]{Schlegel:2019eqc}%
  \BibitemOpen
  \bibfield  {author} {\bibinfo {author} {\bibfnamefont {D.~J.}\ \bibnamefont
  {Schlegel}} \emph {et~al.},\ }\href@noop {} {\  (\bibinfo {year} {2019})},\
  \Eprint {http://arxiv.org/abs/1907.11171} {arXiv:1907.11171 [astro-ph.IM]}
  \BibitemShut {NoStop}%
\bibitem [{\citenamefont {Babusiaux}\ \emph {et~al.}(2019)\citenamefont
  {Babusiaux} \emph {et~al.}}]{MSEScienceTeam:2019bva}%
  \BibitemOpen
  \bibfield  {author} {\bibinfo {author} {\bibfnamefont {C.}~\bibnamefont
  {Babusiaux}} \emph {et~al.} (\bibinfo {collaboration} {MSE Science Team}),\
  }\href@noop {} {\  (\bibinfo {year} {2019})},\ \Eprint
  {http://arxiv.org/abs/1904.04907} {arXiv:1904.04907 [astro-ph.IM]}
  \BibitemShut {NoStop}%
\bibitem [{\citenamefont {Coulton}\ \emph {et~al.}(2020)\citenamefont
  {Coulton}, \citenamefont {Meerburg}, \citenamefont {Baker}, \citenamefont
  {Hotinli}, \citenamefont {Duivenvoorden},\ and\ \citenamefont {van
  Engelen}}]{Coulton:2019odk}%
  \BibitemOpen
  \bibfield  {author} {\bibinfo {author} {\bibfnamefont {W.~R.}\ \bibnamefont
  {Coulton}}, \bibinfo {author} {\bibfnamefont {P.~D.}\ \bibnamefont
  {Meerburg}}, \bibinfo {author} {\bibfnamefont {D.~G.}\ \bibnamefont {Baker}},
  \bibinfo {author} {\bibfnamefont {S.}~\bibnamefont {Hotinli}}, \bibinfo
  {author} {\bibfnamefont {A.~J.}\ \bibnamefont {Duivenvoorden}}, \ and\
  \bibinfo {author} {\bibfnamefont {A.}~\bibnamefont {van Engelen}},\ }\href
  {\doibase 10.1103/PhysRevD.101.123504} {\bibfield  {journal} {\bibinfo
  {journal} {Phys. Rev. D}\ }\textbf {\bibinfo {volume} {101}},\ \bibinfo
  {pages} {123504} (\bibinfo {year} {2020})},\ \Eprint
  {http://arxiv.org/abs/1912.07619} {arXiv:1912.07619 [astro-ph.CO]}
  \BibitemShut {NoStop}%
\bibitem [{\citenamefont {Eisenstein}\ \emph {et~al.}(2007)\citenamefont
  {Eisenstein}, \citenamefont {Seo}, \citenamefont {Sirko},\ and\ \citenamefont
  {Spergel}}]{Eisenstein:2006nk}%
  \BibitemOpen
  \bibfield  {author} {\bibinfo {author} {\bibfnamefont {D.~J.}\ \bibnamefont
  {Eisenstein}}, \bibinfo {author} {\bibfnamefont {H.-j.}\ \bibnamefont {Seo}},
  \bibinfo {author} {\bibfnamefont {E.}~\bibnamefont {Sirko}}, \ and\ \bibinfo
  {author} {\bibfnamefont {D.}~\bibnamefont {Spergel}},\ }\href {\doibase
  10.1086/518712} {\bibfield  {journal} {\bibinfo  {journal} {Astrophys. J.}\
  }\textbf {\bibinfo {volume} {664}},\ \bibinfo {pages} {675} (\bibinfo {year}
  {2007})},\ \Eprint {http://arxiv.org/abs/astro-ph/0604362}
  {arXiv:astro-ph/0604362} \BibitemShut {NoStop}%
\bibitem [{\citenamefont {Coulton}\ \emph {et~al.}(2022)\citenamefont
  {Coulton}, \citenamefont {Villaescusa-Navarro}, \citenamefont {Jamieson},
  \citenamefont {Baldi}, \citenamefont {Jung}, \citenamefont {Karagiannis},
  \citenamefont {Liguori}, \citenamefont {Verde},\ and\ \citenamefont
  {Wandelt}}]{Coulton:2022qbc}%
  \BibitemOpen
  \bibfield  {author} {\bibinfo {author} {\bibfnamefont {W.~R.}\ \bibnamefont
  {Coulton}}, \bibinfo {author} {\bibfnamefont {F.}~\bibnamefont
  {Villaescusa-Navarro}}, \bibinfo {author} {\bibfnamefont {D.}~\bibnamefont
  {Jamieson}}, \bibinfo {author} {\bibfnamefont {M.}~\bibnamefont {Baldi}},
  \bibinfo {author} {\bibfnamefont {G.}~\bibnamefont {Jung}}, \bibinfo {author}
  {\bibfnamefont {D.}~\bibnamefont {Karagiannis}}, \bibinfo {author}
  {\bibfnamefont {M.}~\bibnamefont {Liguori}}, \bibinfo {author} {\bibfnamefont
  {L.}~\bibnamefont {Verde}}, \ and\ \bibinfo {author} {\bibfnamefont {B.~D.}\
  \bibnamefont {Wandelt}},\ }\href@noop {} {\  (\bibinfo {year} {2022})},\
  \Eprint {http://arxiv.org/abs/2206.01619} {arXiv:2206.01619 [astro-ph.CO]}
  \BibitemShut {NoStop}%
\bibitem [{\citenamefont {Leclercq}\ and\ \citenamefont
  {Wandelt}(2014)}]{Leclercq:2014fta}%
  \BibitemOpen
  \bibfield  {author} {\bibinfo {author} {\bibfnamefont {F.}~\bibnamefont
  {Leclercq}}\ and\ \bibinfo {author} {\bibfnamefont {B.}~\bibnamefont
  {Wandelt}},\ }\href {\doibase 10.1017/S1743921314011120} {\bibfield
  {journal} {\bibinfo  {journal} {IAU Symp.}\ }\textbf {\bibinfo {volume}
  {306}},\ \bibinfo {pages} {1} (\bibinfo {year} {2014})},\ \Eprint
  {http://arxiv.org/abs/1410.1546} {arXiv:1410.1546 [astro-ph.CO]} \BibitemShut
  {NoStop}%
\bibitem [{\citenamefont {Cranmer}\ \emph {et~al.}(2020)\citenamefont
  {Cranmer}, \citenamefont {Brehmer},\ and\ \citenamefont
  {Louppe}}]{cranmer2020frontier}%
  \BibitemOpen
  \bibfield  {author} {\bibinfo {author} {\bibfnamefont {K.}~\bibnamefont
  {Cranmer}}, \bibinfo {author} {\bibfnamefont {J.}~\bibnamefont {Brehmer}}, \
  and\ \bibinfo {author} {\bibfnamefont {G.}~\bibnamefont {Louppe}},\
  }\href@noop {} {\bibfield  {journal} {\bibinfo  {journal} {Proceedings of the
  National Academy of Sciences}\ }\textbf {\bibinfo {volume} {117}},\ \bibinfo
  {pages} {30055} (\bibinfo {year} {2020})}\BibitemShut {NoStop}%
\bibitem [{\citenamefont {Alsing}\ \emph {et~al.}(2018)\citenamefont {Alsing},
  \citenamefont {Wandelt},\ and\ \citenamefont {Feeney}}]{Alsing:2018eau}%
  \BibitemOpen
  \bibfield  {author} {\bibinfo {author} {\bibfnamefont {J.}~\bibnamefont
  {Alsing}}, \bibinfo {author} {\bibfnamefont {B.}~\bibnamefont {Wandelt}}, \
  and\ \bibinfo {author} {\bibfnamefont {S.}~\bibnamefont {Feeney}},\ }\href
  {\doibase 10.1093/mnras/sty819} {\bibfield  {journal} {\bibinfo  {journal}
  {Mon. Not. Roy. Astron. Soc.}\ }\textbf {\bibinfo {volume} {477}},\ \bibinfo
  {pages} {2874} (\bibinfo {year} {2018})},\ \Eprint
  {http://arxiv.org/abs/1801.01497} {arXiv:1801.01497 [astro-ph.CO]}
  \BibitemShut {NoStop}%
\bibitem [{\citenamefont {Alsing}\ \emph {et~al.}(2019)\citenamefont {Alsing},
  \citenamefont {Charnock}, \citenamefont {Feeney},\ and\ \citenamefont
  {Wandelt}}]{Alsing:2019xrx}%
  \BibitemOpen
  \bibfield  {author} {\bibinfo {author} {\bibfnamefont {J.}~\bibnamefont
  {Alsing}}, \bibinfo {author} {\bibfnamefont {T.}~\bibnamefont {Charnock}},
  \bibinfo {author} {\bibfnamefont {S.}~\bibnamefont {Feeney}}, \ and\ \bibinfo
  {author} {\bibfnamefont {B.}~\bibnamefont {Wandelt}},\ }\href {\doibase
  10.1093/mnras/stz1960} {\bibfield  {journal} {\bibinfo  {journal} {Mon. Not.
  Roy. Astron. Soc.}\ }\textbf {\bibinfo {volume} {488}},\ \bibinfo {pages}
  {4440} (\bibinfo {year} {2019})},\ \Eprint {http://arxiv.org/abs/1903.00007}
  {arXiv:1903.00007 [astro-ph.CO]} \BibitemShut {NoStop}%
\bibitem [{\citenamefont {Jeffrey}\ and\ \citenamefont
  {Wandelt}(2020)}]{Jeffrey:2020itg}%
  \BibitemOpen
  \bibfield  {author} {\bibinfo {author} {\bibfnamefont {N.}~\bibnamefont
  {Jeffrey}}\ and\ \bibinfo {author} {\bibfnamefont {B.~D.}\ \bibnamefont
  {Wandelt}},\ }in\ \href@noop {} {\emph {\bibinfo {booktitle} {{34th
  Conference on Neural Information Processing Systems}}}}\ (\bibinfo {year}
  {2020})\ \Eprint {http://arxiv.org/abs/2011.05991} {arXiv:2011.05991
  [stat.ML]} \BibitemShut {NoStop}%
\bibitem [{\citenamefont {Miller}\ \emph {et~al.}(2020)\citenamefont {Miller},
  \citenamefont {Cole}, \citenamefont {Louppe},\ and\ \citenamefont
  {Weniger}}]{Miller:2020hua}%
  \BibitemOpen
  \bibfield  {author} {\bibinfo {author} {\bibfnamefont {B.~K.}\ \bibnamefont
  {Miller}}, \bibinfo {author} {\bibfnamefont {A.}~\bibnamefont {Cole}},
  \bibinfo {author} {\bibfnamefont {G.}~\bibnamefont {Louppe}}, \ and\ \bibinfo
  {author} {\bibfnamefont {C.}~\bibnamefont {Weniger}},\ }\href@noop {} {\
  (\bibinfo {year} {2020})},\ \Eprint {http://arxiv.org/abs/2011.13951}
  {arXiv:2011.13951 [astro-ph.IM]} \BibitemShut {NoStop}%
\bibitem [{\citenamefont {Cole}\ \emph {et~al.}(2021)\citenamefont {Cole},
  \citenamefont {Miller}, \citenamefont {Witte}, \citenamefont {Cai},
  \citenamefont {Grootes}, \citenamefont {Nattino},\ and\ \citenamefont
  {Weniger}}]{Cole:2021gwr}%
  \BibitemOpen
  \bibfield  {author} {\bibinfo {author} {\bibfnamefont {A.}~\bibnamefont
  {Cole}}, \bibinfo {author} {\bibfnamefont {B.~K.}\ \bibnamefont {Miller}},
  \bibinfo {author} {\bibfnamefont {S.~J.}\ \bibnamefont {Witte}}, \bibinfo
  {author} {\bibfnamefont {M.~X.}\ \bibnamefont {Cai}}, \bibinfo {author}
  {\bibfnamefont {M.~W.}\ \bibnamefont {Grootes}}, \bibinfo {author}
  {\bibfnamefont {F.}~\bibnamefont {Nattino}}, \ and\ \bibinfo {author}
  {\bibfnamefont {C.}~\bibnamefont {Weniger}},\ }\href@noop {} {\  (\bibinfo
  {year} {2021})},\ \Eprint {http://arxiv.org/abs/2111.08030} {arXiv:2111.08030
  [astro-ph.CO]} \BibitemShut {NoStop}%
\bibitem [{Note6()}]{Note6}%
  \BibitemOpen
  \bibinfo {note} {\protect \url {https://github.com/tsfloss/pyNG}}\BibitemShut
  {NoStop}%
\bibitem [{Note7()}]{Note7}%
  \BibitemOpen
  \bibinfo {note} {\protect \url
  {https://github.com/franciscovillaescusa/Pylians}}\BibitemShut {NoStop}%
\bibitem [{Note8()}]{Note8}%
  \BibitemOpen
  \bibinfo {note} {\protect \url
  {https://github.com/changhoonhahn/pySpectrum}}\BibitemShut {NoStop}%
\bibitem [{\citenamefont {Hartlap}\ \emph {et~al.}(2007)\citenamefont
  {Hartlap}, \citenamefont {Simon},\ and\ \citenamefont
  {Schneider}}]{Hartlap:2006kj}%
  \BibitemOpen
  \bibfield  {author} {\bibinfo {author} {\bibfnamefont {J.}~\bibnamefont
  {Hartlap}}, \bibinfo {author} {\bibfnamefont {P.}~\bibnamefont {Simon}}, \
  and\ \bibinfo {author} {\bibfnamefont {P.}~\bibnamefont {Schneider}},\ }\href
  {\doibase 10.1051/0004-6361:20066170} {\bibfield  {journal} {\bibinfo
  {journal} {Astron. Astrophys.}\ }\textbf {\bibinfo {volume} {464}},\ \bibinfo
  {pages} {399} (\bibinfo {year} {2007})},\ \Eprint
  {http://arxiv.org/abs/astro-ph/0608064} {arXiv:astro-ph/0608064} \BibitemShut
  {NoStop}%
\bibitem [{\citenamefont {Biagetti}\ \emph {et~al.}(2017)\citenamefont
  {Biagetti}, \citenamefont {Lazeyras}, \citenamefont {Baldauf}, \citenamefont
  {Desjacques},\ and\ \citenamefont {Schmidt}}]{Biagetti:2016ywx}%
  \BibitemOpen
  \bibfield  {author} {\bibinfo {author} {\bibfnamefont {M.}~\bibnamefont
  {Biagetti}}, \bibinfo {author} {\bibfnamefont {T.}~\bibnamefont {Lazeyras}},
  \bibinfo {author} {\bibfnamefont {T.}~\bibnamefont {Baldauf}}, \bibinfo
  {author} {\bibfnamefont {V.}~\bibnamefont {Desjacques}}, \ and\ \bibinfo
  {author} {\bibfnamefont {F.}~\bibnamefont {Schmidt}},\ }\href {\doibase
  10.1093/mnras/stx714} {\bibfield  {journal} {\bibinfo  {journal} {Mon. Not.
  Roy. Astron. Soc.}\ }\textbf {\bibinfo {volume} {468}},\ \bibinfo {pages}
  {3277} (\bibinfo {year} {2017})},\ \Eprint {http://arxiv.org/abs/1611.04901}
  {arXiv:1611.04901 [astro-ph.CO]} \BibitemShut {NoStop}%
\bibitem [{\citenamefont {Baldauf}\ \emph {et~al.}(2016)\citenamefont
  {Baldauf}, \citenamefont {Mirbabayi}, \citenamefont {Simonovi\'c},\ and\
  \citenamefont {Zaldarriaga}}]{Baldauf:2016sjb}%
  \BibitemOpen
  \bibfield  {author} {\bibinfo {author} {\bibfnamefont {T.}~\bibnamefont
  {Baldauf}}, \bibinfo {author} {\bibfnamefont {M.}~\bibnamefont {Mirbabayi}},
  \bibinfo {author} {\bibfnamefont {M.}~\bibnamefont {Simonovi\'c}}, \ and\
  \bibinfo {author} {\bibfnamefont {M.}~\bibnamefont {Zaldarriaga}},\
  }\href@noop {} {\  (\bibinfo {year} {2016})},\ \Eprint
  {http://arxiv.org/abs/1602.00674} {arXiv:1602.00674 [astro-ph.CO]}
  \BibitemShut {NoStop}%
\bibitem [{\citenamefont {Tomlinson}\ and\ \citenamefont
  {Jeong}(2022)}]{Tomlinson:2022xud}%
  \BibitemOpen
  \bibfield  {author} {\bibinfo {author} {\bibfnamefont {J.}~\bibnamefont
  {Tomlinson}}\ and\ \bibinfo {author} {\bibfnamefont {D.}~\bibnamefont
  {Jeong}},\ }\href@noop {} {\  (\bibinfo {year} {2022})},\ \Eprint
  {http://arxiv.org/abs/2204.00668} {arXiv:2204.00668 [astro-ph.CO]}
  \BibitemShut {NoStop}%
\bibitem [{\citenamefont {Sailer}\ \emph {et~al.}(2021)\citenamefont {Sailer},
  \citenamefont {Castorina}, \citenamefont {Ferraro},\ and\ \citenamefont
  {White}}]{Sailer:2021yzm}%
  \BibitemOpen
  \bibfield  {author} {\bibinfo {author} {\bibfnamefont {N.}~\bibnamefont
  {Sailer}}, \bibinfo {author} {\bibfnamefont {E.}~\bibnamefont {Castorina}},
  \bibinfo {author} {\bibfnamefont {S.}~\bibnamefont {Ferraro}}, \ and\
  \bibinfo {author} {\bibfnamefont {M.}~\bibnamefont {White}},\ }\href
  {\doibase 10.1088/1475-7516/2021/12/049} {\bibfield  {journal} {\bibinfo
  {journal} {JCAP}\ }\textbf {\bibinfo {volume} {12}},\ \bibinfo {pages} {049}
  (\bibinfo {year} {2021})},\ \Eprint {http://arxiv.org/abs/2106.09713}
  {arXiv:2106.09713 [astro-ph.CO]} \BibitemShut {NoStop}%
\end{thebibliography}%

\end{document}